\documentclass[fleqn,usenatbib]{mnras}

\usepackage{mathptmx}

\usepackage[T1]{fontenc}

\DeclareRobustCommand{\VAN}[3]{#2}
\let\VANthebibliography\thebibliography
\def\thebibliography{\DeclareRobustCommand{\VAN}[3]{##3}\VANthebibliography}

\usepackage[pdftex]{graphicx}	
\usepackage{amsmath}	

\usepackage{amssymb}	
\usepackage{enumitem}
\usepackage[thinc]{esdiff}
\usepackage[flushleft]{threeparttable}

\title[The stings of the scorpion]{Substructure, supernovae, and a time-resolved star formation history for Upper Scorpius}
\author[Briceño-Morales \& Chanamé]{
Geovanny Briceño-Morales,$^{1,2}$\thanks{E-mail: gbriceno@astro.puc.cl}
Julio Chanamé$^{2}$
\\
$^{1}$Centro de Investigaciones Espaciales (CINESPA), Escuela de Física, Universidad de Costa Rica, San Pedro de Montes de Oca 11501, San José, Costa Rica\\
$^{2}$Instituto de Astrofísica, Pontificia Universidad Católica de Chile, Av. Vicuña Mackenna 4860, 782-0436 Macul, Santiago, Chile
}

\date{Accepted 2023 February 15. Received 2023 February 13; in original form 2022 May 03}
\pubyear{2023}

\begin{document}
\label{firstpage}
\pagerange{\pageref{firstpage}--\pageref{lastpage}}
\maketitle

\begin{abstract}
The improved astrometry precision of Gaia-eDR3 allows us to perform a detailed study of the Upper Scorpius OB association and revisit its spatial, kinematic, and age substructure. We achieve this by combining clustering techniques and complementing with age estimations based on Gaia photometry. Our census retrieves 3661 candidate members for Upper Scorpius with contamination $\sim$9\%. We also extract an astrometrically clean sample of 3004 sources with contamination $\sim$6\%. We show that Upper Scorpius can be divided into at least 3 main kinematic groups. We systematically investigate and characterize the Upper Scorpius' internal structure, revealing that at least $\sim 34\%$ of its stellar populations are contained in 7 spatial substructures, with well defined boundaries, kinematics and relative ages, with suggested names: $\pi$ Scorpii (20 $^{\pm2}_{\pm1}$ Myr), $\alpha$ Scorpii (14$^{\pm2}_{\pm1}$ Myr), $\delta$ Scorpii (9$^{\pm2}_{\pm1}$ Myr), $\beta$ Scorpii (8$^{\pm1}_{\pm1}$ Myr), $\omega$ Scorpii (8$^{\pm1}_{\pm1}$ Myr), $\nu$ Scorpii (7$^{\pm1}_{\pm1}$ Myr), after their brightest member, and the well known $\rho$ Ophiuchi (4$^{\pm1}_{\pm1}$ Myr). We ﬁnd a clear correlation in (1) density-age, providing an empirical expansion law to be tested in other associations, and (2) tangential velocity-age, providing constrains on the dynamics of these substructures and the position of potential past triggering events. We estimate the time at which 4 potential supernovae events occurred in Upper Scorpius. Based on these ﬁndings, we tie together previous work on the region and suggest a star formation history with unprecedented temporal resolution.

\end{abstract}







\section{Introduction}\label{1}

OB associations are loose groups of young stars and young stellar objects with prominent OB stars as members. These structures have long been thought to be an intermediate stage between star formation and the less dense young populations of the Galactic field \citep{blaauwO1964}. Hence, observations of nearby OB associations are fundamental for the test and proposal of models ranging from the early stages of stellar evolution to the structure, composition and dynamics of the Galactic populations. In this context, Upper Scorpius (USco) has a great importance as it is thought to be the youngest group of the three conforming the Scorpius-Centaurus OB2 association (Sco-Cen), the nearest to Earth \citeg{deGeusHIshells,deZeeuw1999,PecautSco-cen2016} and one of the nearest places ($\sim$140 pc) that show evidence of active or recent star formation \citep{deZeeuw1997,Preibisch08,rhoOph0myr}. Therefore, in depth efforts have been made in order to provide a systematic characterization of the stellar populations in USco. 

Before the Gaia mission \citep{TheGaiaMission}, only mid to high-mass members were properly identified via astrometric methods, while the expected low mass populations were obtained mainly with photometric and spectroscopic methods \citep{wright2020obreview}, yielding important insights on its features \citeg{preibisch1999,Lodieu2013} but keeping them without a precise description of their astrometric properties. However, pre-Gaia studies were enough to unveil puzzling characteristics of USco like its broad age dispersion \citep{Pecaut2012} and complicated structure \citep{PecautSco-cen2016,wrigth}, leading to the proposal of multiple scenarios to explain this \citeg{Feiden2016,trevor2019,Sullivan2021}. 

Several studies have been done to update the USco low mass populations membership revealed by the Gaia data \citeg{cookdr1, wolkinsondr1,Damiani2019,Luhman2020,Luhman2022}, as it provides unprecedented completeness and precision in its five parameter (sky position, parallax and proper motion) astrometrical solution. However, until recently, no one had been able to characterize in a systematic way the internal structure of USco as revealed by this data, mostly due to the apparently high level of mixing seen on its internal stellar populations that does not seem to show any pattern in the spatial nor kinematic spaces \citep{Damiani2019}. The latter was partially solved by the recent work of \cite{kerr}, where the authors use the HDBSCAN clustering algorithm to succesfully find and characterize young substructures up to a distance of $\sim$333 pc around the Sun, including substructure in Sco-Cen and USco. However, even though the work of \cite{kerr} provides an important first insight on the properties of USco, this particular implementation of HDBSCAN is affected by the definition of the clustering space when resolving the very local spatial and kinematic substructure of USco, as we will discuss in Section (\ref{3.2.1}). This is further affected by the lower precision of the Gaia DR2 with respect to the Gaia eDR3 catalogue, i.e., the typical errors in Gaia DR2 \citep{Gaiadr2} are larger by a factor of $\sim$ 3 in proper motion and 30-50\% in parallax. More recently, \cite{Squicciarini} found kinematic evidence of substructure in USco by tracing back the positions of USco candidate members. Via this approach, the authors were able to recover substructures that visually resemble some of the retrieved by \cite{kerr}. However, a robust validation and in depth characterization of these substructures, if validated, has not been performed yet. Another open problem in USco is that age estimations yield a rather broad age dispersion \citeg{PecautSco-cen2016,Luhman2022} and, until now, there was no clear pattern in the ages of the local USco substructures found with Gaia. We will show that this age problem has a natural solution if a multidimensional analysis is performed, systematic effects are considered when extracting isochronal ages and the massive membership and concrete past supernovae events in USco, if any, are investigated. In short, we aim to contribute with an in depth and systematic characterization of the internal structure of USco to extract new information about this important region, all based solely on the Gaia eDR3 catalogue \citep{gaiaedr3}.

This work is organized as follows: in Section \ref{2}, we summarize the limits that contain our main data set of the USco region, define some useful notation and describe the subsample of massive stars. In Section \ref{3} we describe: the Kinematic Analysis, i.e., the implementation of the Gaussian Mixture (GM) fit \citep{scikit-learn} to identify kinematic groups (KGs) in velocity space with a Bayesian approach, the Spatial Analysis, where we use OPTICS \citep{OPTICS} to directly see the internal structure of USco in the (X,Y,Z) Galactic Heliocentric Coordinates and the Time Analysis, where we investigate the input parameters of the present-day mass function in order to estimate the number and time of supernovae events occurred in USco.
In Section \ref{4}, we show that this approach from multiple directions proves to be useful because it gives new detailed and relevant information about the substructure, kinematics, dynamics, age, star formation history of USco and yields new constrains in the fundamental properties of its massive members. In Section \ref{5} we summarize and discuss our results in relation with the age and past supernovae events in USco to suggest a star formation scenario with unprecedented temporal resolution.

\section{Data and notation}\label{2}

\subsection{Gaia}\label{2.1}

We download all Gaia eDR3 sources contained within the sky region defined by \cite{deZeeuw1997}, in Galactic coordinates ($\ell,b$), and closer than 200 pc according to the (inverse of) eDR3 parallax ($\varpi$). That is:

\begin{gather}
    343^\circ < \ell <  360^\circ \\
    10^\circ < b < 30^\circ \\
    \varpi > 5 \: mas \;
\end{gather}

The upper distance of $\sim$ 200 pc is taken following \cite{Damiani2019} (its Sect.5.3). Distance estimation by inversion of the parallax constitutes a good approximation at the distance of our analysis \citep{Luri2018}. The resulting data set contains 51689 sources and we will refer to it as the Raw Gaia-eDR3 set. We aim to compile a complete sample of candidate members from this set, but the nature of the analysis in this work requires a small-error subset. Hence, we apply the following quality criteria \citep{gaiaedr3astrometricsolution} on the Raw Gaia-eDR3 set's parameters:
 \begin{gather}
    \varpi/\mathrm{error}_\varpi > 10 \label{eqqualityparallax}\\ 
    \mathrm{RUWE} < 1.4 \label{eqqualityruwe} \; 
 \end{gather}
where the upper limit in RUWE provides a highly clean five-parameter solution \citep{Lindegren}. The resulting data set is composed by 17535 sources and we refer to it as Gaia-eDR3 set.

Furthermore, we refer to the sample from the Gaia-eDR3 set with available radial velocities (V$_r$) as the Gaia-eDR3$_{\mathrm{V_r}}$ sample (2310 sources). The distinction is necessary as only the latter sample can be fully described in the (U,V,W) space. To describe the kinematic properties of sources with no V$_r$ measurements, we use the tangential velocity offset ($\boldsymbol{\mathrm{V}}_{\mathrm{T}}*$) obtained after correcting perspective effects from the tangential velocities ($\boldsymbol{\mathrm {V}}_{\mathrm{T}}$) (see Appendix \ref{A}), which is necessary due to the large sky area under study ($\sim$340 sq.deg.). We work in the Galactic Cartesian heliocentric positions (X,Y,Z), with axes pointing to:\;

\begin{itemize}
  \let\labelitemi\labelitemii
    \item X (pc): Galactic center
    \item Y (pc): direction\:of\:Galactic\:rotation
    \item Z (pc): North\:Galactic\:Pole
\end{itemize}

and the respective Galactic Cartesian velocities (U,V,W) km s$^{-1}$. 
We also denote the other main kinematic variables as:

\begin{itemize}
  \let\labelitemi\labelitemii
    \item $\mu_i$ (mas yr$^{-1}$): i component of the proper\:motion
    \item V$_i$ (km s$^{-1}$): i component of the tangential\:velocity
    \item V$_{i \mathrm{exp}}$ (km s$^{-1}$): i component of the expected\:tangential\:velocity
\end{itemize}

\subsection{Massive stars}\label{2.2}

We also investigate the relation between the most massive stars from USco and its internal substructure. In order to do this, we compile a sample with the turn-off stars in USco from \cite{PecautSco-cen2016} (its Table 10) plus Antares ($\alpha$ Sco$_\star$)\footnote{To avoid confusion in further sections, we will refer to these stars by their approved names (see \url{https://www.iau.org/public/themes/naming_stars/}.), or include the sub index "$\star$" in their Bayer designation.} and we refer to this as the Turn-off sample. There are four stars with no $\varpi$ or $\mu$ measurements from Gaia: Antares, Dschubba ($\delta$ Sco$_\star$), Paikauhale ($\tau$ Sco$_\star$) and Alniyat ($\sigma$ Sco$_\star$). We use astrometry from HIPPARCOS \citep{Hipparcos2007} for these stars. In the particular case of Alniyat, we use the improved distance from \cite{sigmaScoNorth2007} and estimate its $\boldsymbol{\mathrm V}_{\mathrm T}$ based on it.

\section{Methods}\label{3}

It is convenient to divide our analysis in three main branches clearly distinguished by the spaces where they are developed into; the Kinematic, Spatial and Time analysis. Later, we will see how the information obtained from each space merges to tie and constrain our knowledge about USco.

\subsection{Kinematic Analysis}\label{3.1}

By taking advantage of the high precision astrometry of the Gaia-eDR3 data, we perform a variation of the Convergent Point method, as implemented by \cite{perryman1998}, that intends to allow us to distinguish kinematic substructures in USco, if any, and their respective spatial distributions and ages. We will refer to the application of this method as the Kinematic Analysis and we explain it and summarize it as follows.


1. The existence of an OB association and its presence in the data implies two fundamental facts: OB stars must be present and they must have similar kinematic properties. To assess this, we search for OB stars in our Raw Gaia-eDR3 set. This can be done by performing a nearest neighbor sky crossmatch of the OB stars catalog from \cite{obstarsgontcharov} with this set. We also aim to obtain a more complete sample by transforming the observed color index to the intrinsic spectral type of the stars, we opt to do this making use of the intrinsic Gaia colors from \cite{Luhman2022} and near-IR color-T$_{\mathrm{eff}}$ estimates for O9V-M9V MS stars from \cite{Pecaut2013SpT} (its Table 5)\footnote{An updated version of the table can be found at https://www.pas.rochester.edu/$\sim$emamajek/EEM\_dwarf\_UBVIJHK\_colors\_Teff.txt}. For the latter, we crossmatch the set with the near-IR 2MASS catalog \citep{2MASS} and select the OB stars according to the corrections of \cite{Pecaut2013SpT}. We complement our selection with the stars contained in the Turn-Off sample (see Section \ref{2.2}) and double check with the catalog from \cite{skiff2009}. With this method, we expect to obtain a complete sample of stars with spectral type (SpT) $<$ B0. We refer to the final sample of OB stars candidates (plus Antares) as the OB sample. 

2. To identify the underlying stellar populations related to a group of clustered OB stars, if any, in a given kinematic space, we fit the velocity distribution of the Gaia-eDR3 set with a GM model. For the Gaia-eDR3$_{\mathrm{V_r}}$ sample, we are able to apply the GM fit \citep{scikit-learn} to the data in the (U,V,W) space. The number of components of the fit is chosen according to the Bayesian Information Criterion (BIC) (see Appendix \ref{C}). If none or only a few OB stars have available V$_\mathrm{r}$ measurements, the component that corresponds to the OB association, if any, can be recognized by comparing with the respective positions of the OB stars in the $\boldsymbol{\mathrm V}_{\mathrm T}$ space, so that we choose the one (or the ones) that better overlap with the OB sample's velocity distribution and study its physical properties. The mean (U,V,W) values of the selected component can be identified with the true barycenter motion of the association (see Appendix \ref{A}). We refer to this step in the Kinematic Analysis as the First GM fit and Figure \ref{figR.1} shows the result of applying it to the OB and Gaia-eDR3$_{\mathrm{V_r}}$ samples.

3. Formally, the stars have to be clustered in the (U,V,W) velocity space, but most ($\sim88\%$) of the data in the (Raw) Gaia-eDR3 data do not have available V$_{\mathrm{r}}$ measurements, so we have to deal in a different way with these sources. We do this by applying the GM fit a second time in the $\boldsymbol{\mathrm{V}_{\mathrm{T}}}*$ space, which is calculated via the true barycenter motion retrieved from the First GM fit (see Appendix \ref{A}). We run this Second GM fit in the $\boldsymbol{\mathrm{V}}_{\mathrm{T}}*$ space for the (Raw) Gaia-eDR3 set with the the number of components given by the BIC. In this space, the association's members are expected to be near the origin, hence, we select the component with mean $\boldsymbol{\mathrm{V}}_{\mathrm{T}}*$ $\sim$(0,0). Moreover, we can calculate the posterior probabilities for each star to belong to the selected component, hence, we are able to select a confidence region, depending on the desired precision, to perform the rest of the analysis. We will refer to the final confidence region of the selected component from the (Raw) Gaia-eDR3 set as the (Raw) USco KG, which constitute the candidate members' census of the OB association.

4. We can further look for kinematic substructures that could not be obtained in the Second GM fit on the $\boldsymbol{\mathrm{V}}_{\mathrm{T}}*$ space\footnote{This may happen if the true over densities are too local or if they are too close from each other.}. The procedure is the same as before with the only difference that the fit is performed on the USco KG rather than the Gaia-eDR3 set. We rely on the BIC criteria for this new fit to say if the selected component is better modeled with a new mixture of components or not, and interpret the results as a suggestion of the absence or presence of kinematic substructures within the USco KG.

This approach overcomes the necessity of relying on previous results about the presence and kinematic properties of an OB association and allows us to look for kinematic substructures instead of assuming only one KG. Furthermore, the method keeps the advantage of not having to make any assumptions about the spatial distribution of the OB association, so that not only spatially clustered stars are taken as members but also spatially diffuse candidates are included \citep{perryman1998}. This is an advantage when dealing with OB associations as they are not necessarily spatially dense populations. 

\subsubsection{Contamination and mixing}\label{3.1.1}

The GM fit also allows us to estimate not only the level of contamination from field stars but also the internal mixing percentile of the KGs, if present, based on the fitted distributions associated to each one of them. The correspondent posterior probabilities can be calculated for each star with the Bayes theorem:
\begin{equation}
    \mathrm{P}(\mathrm{KG}_\mathrm{i}|\mathrm{V}_\mathrm{T}*) = \frac{\mathrm{P}(\mathrm{V}_\mathrm{T}*|\mathrm{KG}_\mathrm{i})\cdot\mathrm{P}(\mathrm{KG}_\mathrm{i})}{\mathrm{P}({\mathrm{V}_{\mathrm{T}*})}}
    \label{BayesTheorem}
\end{equation}
Here $\mathrm{P}(\mathrm{KG}_\mathrm{i}|\mathrm{V}_\mathrm{T}*)$ is the posterior probability for a star to belong to the i-th component fitted, i.e., the i-th KG, given a position in the $\boldsymbol{\mathrm{V}_\mathrm{T}}*$ space. The likelihood $\mathrm{P}(\mathrm{V}_\mathrm{T}*|\mathrm{KG}_\mathrm{i})$, the prior $\mathrm{P}(\mathrm{KG}_\mathrm{i})$ and the normalization constant  $\mathrm{P}(\mathrm{V}_\mathrm{T}*)$ are obtained with the sci-kit learn GM fit implementation \citep{scikit-learn}. 
The percentiles representing how many stars from the i-th KG are in a given region of the $\boldsymbol{\mathrm{V}_\mathrm{T}}*$ space, can be calculated as
\begin{equation}
    \%_{mix} = \frac{\mathrm{P}(\mathrm{KG}_\mathrm{i}|\mathrm{V}_\mathrm{T}*)}{\Sigma_\mathrm{i} \mathrm{P}(\mathrm{KG}_\mathrm{i}|\mathrm{V}_\mathrm{T}*)}\label{eqmix}
\end{equation}
This is the mixing between different components or KGs, which may refer to the Galactic field, the association's sources or, if present, the internal mixing of the association's kinematic substructures. 
However, this implementation has as major limitations that it is based on the strong assumption of the data to be normal-distributed, and that it does not take into account the errors involved. In particular, the latter assumption may lead to systematic biases if the typical errors are in the same order of magnitude than 1$\sigma$ of the smallest component fitted by the model \citep{Astrostatistic2014izvbezick}. 

As we will show, the Kinematic Analysis proves to be useful when separating USco from the Galactic field, and allows for further spatial substructure to become apparent, but fails to resolve it cleanly. Hence, we need to complement with a clustering algorithm which is able to resolve this substructure, while dealing with noise, and making no prior assumptions about the distribution or density of a cluster, if any, in the space under analysis. We describe such method in the following section.

\subsection{Spatial analysis}\label{3.2}

The OPTICS algorithm \citep{scikit-learn} is a density-based clustering technique that allows to directly see the intrinsic density structure of any n-dimensional data set via its Reachability (R) plot \citep{OPTICS}. We opt to implement OPTICS in our analysis because of its great advantage of not assuming a number of present clusters or any prior distribution on the data and because of its ability to extract clusters with different densities. Here, we introduce the main concepts involved and a guide for the interpretation of the R plot. For a thorough exposition and a recent implementation in the $\rho$ Ophiuchi region see \cite{OPTICS} and \cite{Canovas2019}, respectively. 

\subsubsection{Fundamental concepts}\label{3.2.1}

The local density of an n-dimensional data set can be defined as the number of points inside a given radius $\epsilon$. A core point $o$, then, is defined as a local point that minimizes $\epsilon$ for a given minimum number of points ($MinPts$)\footnote{Note that core points are not necessarily unique.}. \cite{OPTICS} defines as \textit{directly density-reachable} with respect to $o$ every point inside $\epsilon$ and \textit{density-reachable} every point in the transitive hull of direct density reachability limited by the $MinPts$ parameter, i.e., a point is density-reachable if its directly density-reachable from another directly density-reachable point from $o$ and so on, where the $MinPts$ parameter limits that hull. Furthermore, two points $p$ and $q$ are density-connected if both are density-reachable from $o$. Any two points $p$ and $q$ that belong to a cluster must be both density-reachable from $o$ and density-connected with each other, if not, the points are classified as noise or non-clustered points. OPTICS makes this analysis for each point; identifies the core points and obtains $\epsilon$ for each point with respect to the nearest core points making it possible to order the data according to this. The information about the density structure of the data is then contained in the Order vs $\epsilon$ space. Although OPTICS does not consider uncertainties in its procedure, it is useful to consider the order of magnitude of the typical uncertainty in the $\epsilon$ space and compare with the $\epsilon$ values in the R plot to place a quantitative criteria for the statistical significance of the cluster retrieved. We consider this together with the results from the Kinematic Analysis to choose the $\epsilon$ space to be the Galactic Heliocentric Coordinates (X,Y,Z). Then, when applied to the (X,Y,Z) space, OPTICS yields information about the spatial structure of the region and the potential clusters, if any, which can be extracted from the R plot. 

\subsubsection{The Reachability Plot}\label{3.2.2}

The R plot shows the $order$ vs $\epsilon$ space and can be interpreted as follows: clusters can be identified as valleys, and a lower $\epsilon$ for the core points suggest a more dense cluster. The size of the cluster (number of sources) is identified with the size in the $order$ dimension. Even tough the global structure of the R Plot is rather insensitive to the input parameters (see \cite{OPTICS}, Section 4.1), its local shape has dependency on certain parameters; as the number of density-reachable points is proportional to $MinPts$, a large value will soften the curve because more points will be classified as part of a cluster. On the contrary, a low value will produce spurious results, i.e., clusters with size smaller than it is possible to resolve with the data given the typical uncertainty, and will classify as noise points that could be part of substructure. A broad range of $MinPts$ can be chosen without changing the global structure seen in the R Plot and an optimal value can be chosen via heuristic methods or simple inspection, with minimum effects on the cluster identification. Even tough the interpretation of the R plot is very straightforward, the selection of the boundaries of a cluster is still rather subjective. The limits, while not the shape of the plot, depend on the $\xi$ parameter, which defines the steepness at which the cluster starts and ends with and is set depending of the resolution required. Lower values of $\xi$ extract broader structures while higher values restrict the cluster classification to higher steepness in the cluster limits. That is, given a local valley in the R plot, the higher we choose $\xi$ the more conservative we are about the accepted gradient of $\epsilon$ with respect to $order$ at the limit of the valley.

\subsection{Time Analysis}\label{3.3}

In this section we describe how we estimate the number of supernovae (SNe) events occurred for a given substructure. In order to do this, we adopt the \cite{imfKroupa2001} canonical initial mass function (IMF) and the stellar evolution grids with rotation from \cite{ekstromgrids}. The IMF has the form
\begin{equation}
    \diff {\mathrm N}{\mathrm M} = k \cdot \mathrm M^{-\Gamma}
\end{equation}
Where N is the number of stars or systems, M the respective mass and $\Gamma=2.3$. The normalization constant k is obtained with:
\begin{equation}
   k = \frac{\mathrm N }{ \int^{\mathrm M_{up}}_{\mathrm M_{\mathrm{low}}} \mathrm M^{-\Gamma} \mathrm {dM} }
\end{equation}

Where $\mathrm M_{\mathrm{low}}$ is the lower bound in the range of masses spanned by the Turn-Off sample, i.e., approximately the mass of a B2 star, which we assign to be 7.3 M$_\odot$, based on the color-mass estimates from \cite{Pecaut2013SpT} and the measured mass of $\beta^2$ Sco$_\star$ of 7.3$\pm$1 M$_\odot$. Since we expect the minimun mass of supernovae progenitors to be $\sim$ 8 M$_\odot$ \citep{SupernovaMassLimit}, integration in the mass range covered by the Turn-off sample should include all of these progenitors present in USco. M$_{\mathrm {up}}$ is the mass of the most massive member in the given stellar population. Assuming that a well behaved function describes the mass-lifetime relation for H and He-burning stages, we estimate the stellar upper mass, H-M$_{\mathrm {up}}$ and He-M$_{\mathrm {up}}$, at a given age of the parent structure by fitting a spline function (Spl) to this relation from \cite{ekstromgrids}. This way, we are able to compare the available mass measurements of the most massive stars with the predicted upper limit for the structure they belong to. Furthermore, with this fit we are able to obtain the progenitor's mass M$_\tau$ of a SN event occurred $\tau$ Myr ago (this is He-M$_{\mathrm {up}}$($\tau$)), by obtaining the roots of the Spl via the Newton method. We do not consider pre-MS times for this massive stars nor C-burning or latter stages, as they are at least 2 and 3 orders of magnitude lower than the H and He-burning stages, respectively \citep{pmssiess,ekstromgrids}. The number of SN events (N$_{\mathrm {SN}}$) that took place within a stellar population $\tau$ Myr ago is:

\begin{equation}
    \mathrm{N_{SN}} = \int^{\mathrm M_{\tau}}_{\mathrm M_{\mathrm{low}}} k \cdot \mathrm M^{-\gamma} \mathrm{dM}
\end{equation}

N$_{\mathrm{SN}}$ depends on N, M$_{\mathrm{up}}$ and M$_\tau$. When performing this Analysis, we will show that N is sufficiently well constrained. In several cases the observed M$_{\mathrm{up}}$ values lack precision such that no conclusions could be drawn if only relying on them, however, we also show that approximations can be made to obtain results with statistical relevance and high certainty. Uncertainty in M$_\tau$ directly depends on the model that we are assuming. Additionally, N$_{\mathrm{SN}}$ will also systematically depend on the age of the stellar population under analysis, hence, we report and consider the values related to the upper and lower bounds of the ages estimated for each of the substructures.

\begin{figure}
    \centering
    \includegraphics[width=0.45\textwidth]{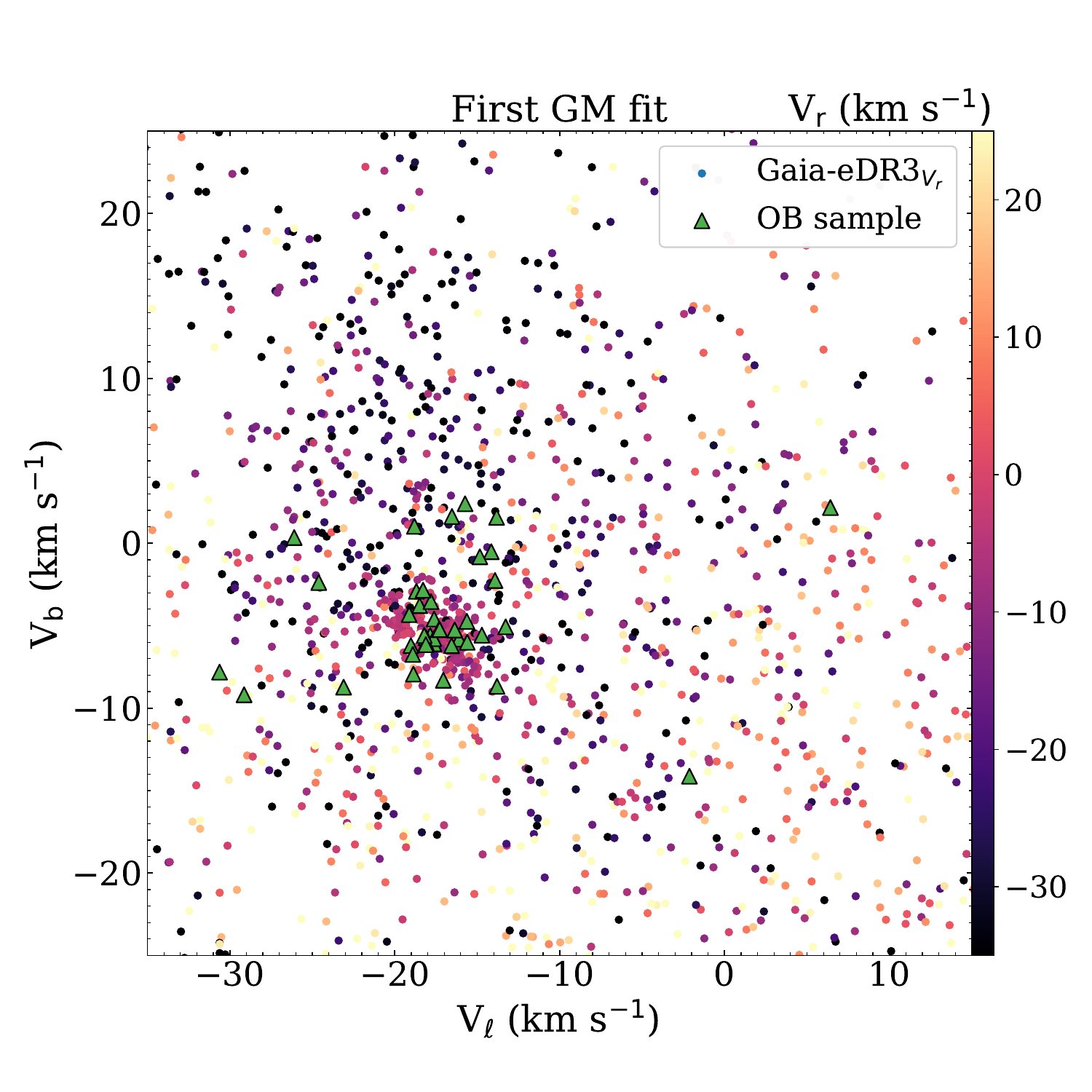}
    \caption{$\boldsymbol{\mathrm{V}_\mathrm{T}}$ space for the Gaia-eDR3$_{\mathrm{V}_\mathrm{r}}$  Sample, where the first GM fit is performed and the OB stars sample. Each cross denote the mean values of a component produced by the GM fit in the (U,V,W) space. Information about V$_\mathrm{r}$ can be extracted from the color map, the min and max color is assigned to sources outside the color bar's interval. The over density of OB stars gives important information of both the presence and kinematic properties of USco as an OB association. The barycenter motion of USco is chosen as the mean position in the (U,V,W) space of the GM component that better overlaps with the OB stars in the $\boldsymbol{\mathrm{V}_\mathrm{T}}$ space.}
    \label{figR.1}
\end{figure}

\begin{figure*}
    \centering
    \includegraphics[width=17cm]{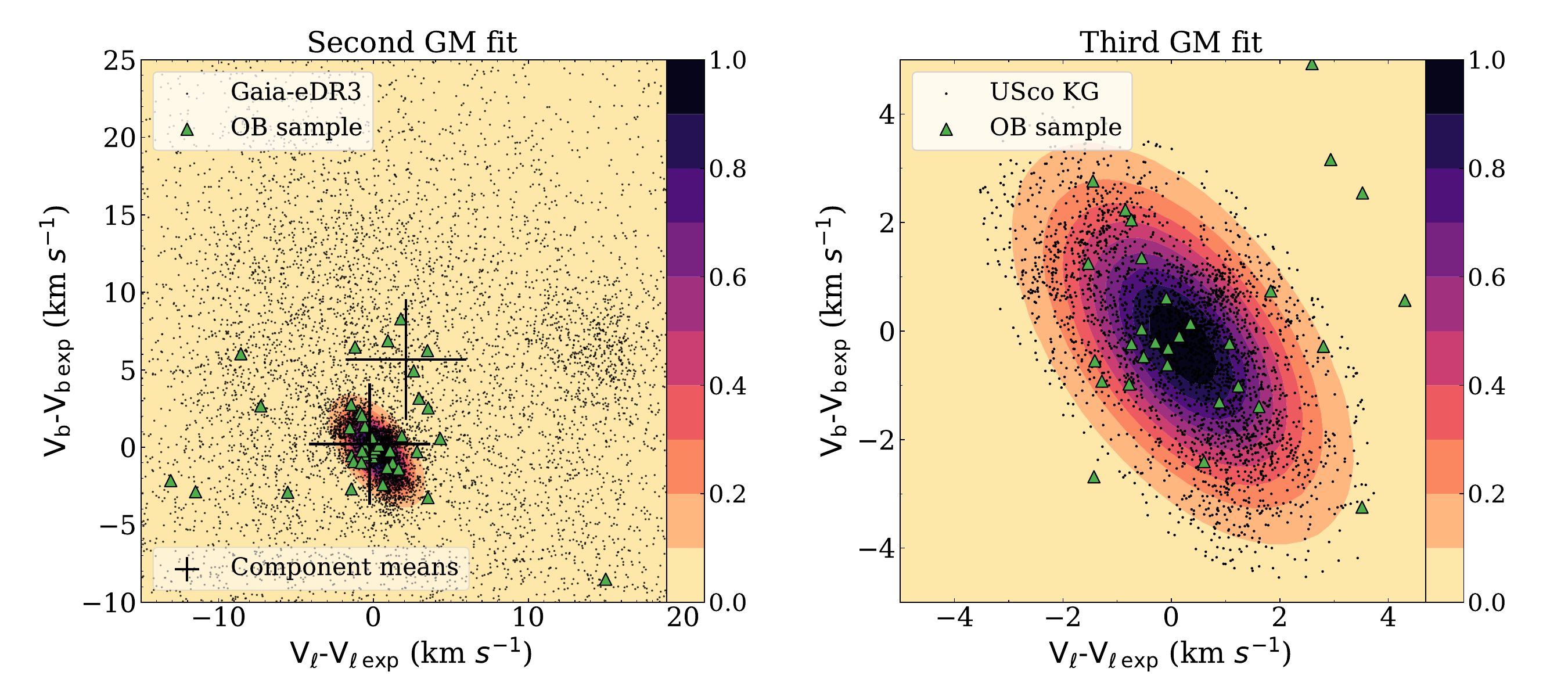}
    \caption{Left: $\boldsymbol{\mathrm{V}}_\mathrm{T}*$ space for the Gaia-eDR3  set.  Here the USco's kinematic structure appears as a notorious over density near the origin. The USco KG is composed by the sources within the 3$\sigma$ regime of the component associated with this over density. The BIC suggest 4 components to fit the Gaia-eDR3 set. One of this components is related to the Galactic field stars (around $\sim$ 15, $\sim$ 7 km s$^-1$).
    Right: $\boldsymbol{\mathrm{V}}_\mathrm{T}*$ for the USco KG. In the Third GM fit, the BIC is minimized if 3 components are chosen, suggesting the presence of 3 main distinct KGs in USco. $\boldsymbol{\mathrm{V}}_\mathrm{T}*$ space is estimated based on the (U,V,W)$_bar$ values obtained in the First GM.
    The color map shows the posterior probability (min-max normalized) produced by the GM fit from the Gaia-eDR3 set and the USco KG in the Second and Third GM fit respectively.}
    \label{figR.2}
\end{figure*}

\section{Results}\label{4}

\subsection{New members in USco}\label{4.1}

Following the Kinematic Analysis, we recuperate 0 O stars and 31 B stars from \cite{obstarsgontcharov} in the Raw Gaia-eDR3 set. We also obtained 17 photometric B stars in the Raw Gaia-eDR3 sample and by doing a crossmatch with the 2MASS catalog with a nearest-neighbor crossmatch with radius of 1 arcsec, we retrieved 23 B stars. After dealing with duplicates, we compile a set of 52 OB stars. We compared the latter sample with the Turn-Off sample (see Sec. \ref{2.2}) and found that $\nu$ Sco$_\star$, $\beta^2$ Sco$_\star$, $\omega^1$ Sco$_\star$, $\pi$ Sco$_\star$, 1 Sco$_\star$, 13 Sco$_\star$ and $\rho$ Sco$_\star$ are contained in our OB sample; $\tau$ Sco$_\star$, $\sigma$ Sco$_\star$ and $\beta^1$ Sco$_\star$ have no $\varpi$ measurements in Gaia and $\delta$ Sco$_\star$ and $\alpha$ Sco$_\star$ have no Gaia counterpart. We add the 5 missing stars from the Turn-Off sample to the OB sample\footnote{Note that $\alpha$ Sco$_\star$ is actually an evolved M type star.}, and obtain a total of 57 OB stars. Finally, we double check with \cite{skiff2009} and discard 9 sources as A stars. The final OB sample contains 47 stars. For $\sigma$ Sco$_\star$, $\beta^1$ Sco$_\star$, $\delta$ Sco$_\star$ and $\alpha$ Sco$_\star$ we use astrometry from Hipparcos \citep{Hipparcos2007}.

We applied the GM fit\footnote{Each GM fit performed in this work converged with a threshold of at least 0.001 km s$^{-1}$} to the Gaia-eDR3$_{\mathrm{V}_\mathrm{r}}$ sample in the (U,V,W) space, this yielded an optimal 3-component mixture, according to the BIC (see Appendix \ref{C}). Figure \ref{figR.1} shows the positions in the $\boldsymbol{V}_\mathrm{T}$ plane in Galactic coordinates (as none or only a few sources have V$_\mathrm{r}$ in Gaia) for the Gaia-eDR3$_{\mathrm{V}_\mathrm{r}}$ sample, the  OB sample and the 4 nearest component means to the OB stars, naturally, our selected component is the one with mean (V$_{\ell} \sim$ --17.3, V$_{\mathrm{b}} \sim$ --5.4) km s$^{-1}$). Hence, we define (U$_0 =$ --5.5$\pm$2.7, V$_0 =$ --16.8$\pm$1.7, W$_0 =$ --6.6$\pm$1.9 km s$^{-1}$) as the true barycenter motion (U, V, W)$_{\mathrm{bar}}$ of USco, which is consistent with previous estimates from Gaia-DR2 (e.g, U, V, W $\sim$ --5, --16, --7 km s$^{-1}$ by \cite{Luhman2018}) and Gaia-DR1 (e.g, U, V, W $\sim$ --6.16, --16,89, --7.05 km s$^{-1}$ by \cite{wrigth}). From the obtained (U, V, W)$_{\mathrm{bar}}$, we calculate $\boldsymbol{\mathrm{V}_\mathrm{T}*}$ for each source in the Gaia-eDR3 set. We applied a Second GM fit in the ($\boldsymbol{\mathrm{V}_\mathrm{T}*}$) space for the Gaia-eDR3 set (left panel in Figure \ref{figR.2}) and selected 4 as the optimal component number according to the BIC. In this space, the component selected is the nearest one to ($\boldsymbol{\mathrm{V}_\mathrm{T}*}$) $=$ (0, 0) km s$^{-1}$) and we chose a confidence region of 3$\sigma$ (3004 sources) from this component as our USco KG (see Section \ref{3.1}). The USco KG represents the entire census of USco and is shown in the right panel of Figure \ref{figR.2}. The estimated contamination in the USco KG is $\sim$ 6\% according to Eq.\eqref{eqmix}. 

We perform the analog procedure for the Raw Gaia-eDR3 set based on the OB sample already described. The retrieved sample is composed by 3661 sources with estimated contamination extent $\sim$ 9\% according to Eq.\eqref{eqmix}. We found that applying at this point the cuts in Eqs. \ref{eqqualityparallax}, \ref{eqqualityruwe} to the this sample, produces roughly the same data set as the USco KG extracted from the Gaia-eDR3 set. 

\subsubsection{Hints of older populations}\label{4.1.1}

Some B stars in Figure \ref{figR.1} seem to be associated with a broader apparent over density of negative radial velocity sources. As counterpart, left panel of Figure \ref{figR.2} shows two out of four components yielded by the second GM fit (other two fall outside the limits of the figure), as discussed, one of the components is related to USco, but the other component of the second GM fit with its mean centered about V$_{\ell}$ - V$_{\ell \: \mathrm{exp}}$ $\sim -2$, V$_{\mathrm{b}}$ - V$_{\mathrm{b\:exp}}$ $\sim$ 6) km s$^{-1}$ is also related to a notable over density of B stars. A possible explanation for this over densities and their B star counterparts, could be the presence of an older population of stars with barely recognizable kinematic properties. For older populations, we expect their kinematic coordinates to spread, and migrate to merge with those of the Galactic field, making it hard to clearly distinguish them from it. This is consistent with the position and qualitative properties of this component of the second GM fit, however, an in depth analysis of this result is beyond the scope of this work. On the other hand, we do not found potential physical relevance in any other component extracted from the First or Second GM fits.

\subsubsection{Membership validation}\label{4.1.2}

We compile a set of of USco members based on the previous census on Sco-Cen by \cite{Damiani2019,Luhman2022}. These census rely on different selection criteria, data releases from Gaia and report the most detailed characterization of USco with the Gaia data, providing together an ideal sample to validate the membership of the Raw USco KG. After extracting duplicates we obtain 3811 sources. We compare our census with the latter sample. This implies that 338 new sources, with respect to these works, are contained in our work. However, CMD positions for this members suggest they are most likely contamination, coinciding with the percentage estimated in Section \ref{4.1}.

\subsubsection{Binaries}\label{4.1.3} 

We note that the large majority of the inferior quality sources in the Raw USco KG are classified as such due to the RUWE threshold in Eq. \eqref{eqqualityruwe}, which is know to be related with unresolved pairs \citep{gaiaedr3astrometricsolution}. We further investigate this by comparing the Raw USco KG and the USco KG samples in the CMD and the {\fontfamily{cmtt}\selectfont ipd\_gof\_harmonic\_amplitude} and {\fontfamily{cmtt}\selectfont ipd\_frac\_multi\_peak} parameters, where the latter two yield information about the partially resolved and resolved binaries, respectively \citep{gaiaedr3astrometricsolution}.

\begin{figure}
    \centering
    \includegraphics[width=8.5cm]{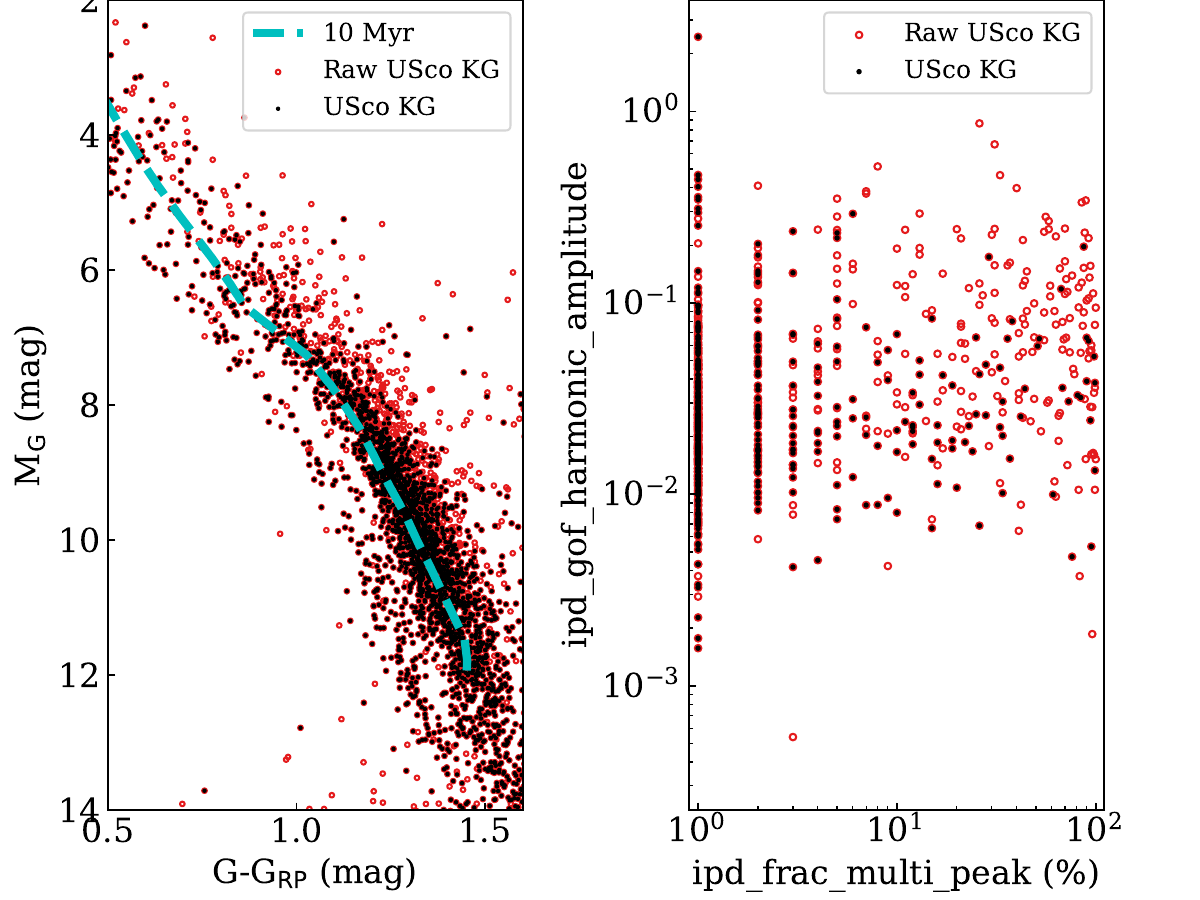}
    \caption{CMD with a 10 Myr isochrone (left) and {\fontfamily{cmtt}\selectfont ipd\_frac\_multi\_peak} vs {\fontfamily{cmtt}\selectfont ipd\_gof\_harmonic\_amplitude} Gaia parameters for the Raw USco KG and the USco KG. Black dots are the inferior quality sources the where left out from the USco KG mostly due to the RUWE criterion, which is well know to account for unresolved pairs \citep{gaiaedr3astrometricsolution}. About half ($\sim$ 500) of this inferior quality sources have photometry consistent with being pre-MS stars. These inferior quality sources also tend to have high values of {\fontfamily{cmtt}\selectfont ipd\_frac\_multi\_peak} and {\fontfamily{cmtt}\selectfont ipd\_gof\_harmonic\_amplitude}, which are sensitive to resolved and partially resolved binaries respectively \citep{gaiaedr3astrometricsolution}.} \label{figR.3}
\end{figure}

In the left panel of Figure \ref{figR.3} we show the CMD for the Raw USco KG (black) and USco KG (red) with a 10 Myr PARSEC isochrone. We note form Figure \ref{figR.3} that about half ($\sim$ 500) of these inferior quality sources have pre-MS positions in the CMD, and hence are true USco members, however, their astrometric quality is diminished as they are probably unresolved, partially resolved or very close binaries. We further support the latter on their trend to have high values of the {\fontfamily{cmtt}\selectfont ipd\_gof\_harmonic\_amplitude} and {\fontfamily{cmtt}\selectfont ipd\_frac\_multi\_peak} parameters, which are related to partially resolved and close resolved binaries respectively \citep{gaiaedr3astrometricsolution}. The presence of this large fraction of close binaries may yield important information on the dynamics of USco, but a further investigation on the properties of these sources escapes the aims of this work.

\begin{figure}
    \centering
    \includegraphics[width=8cm]{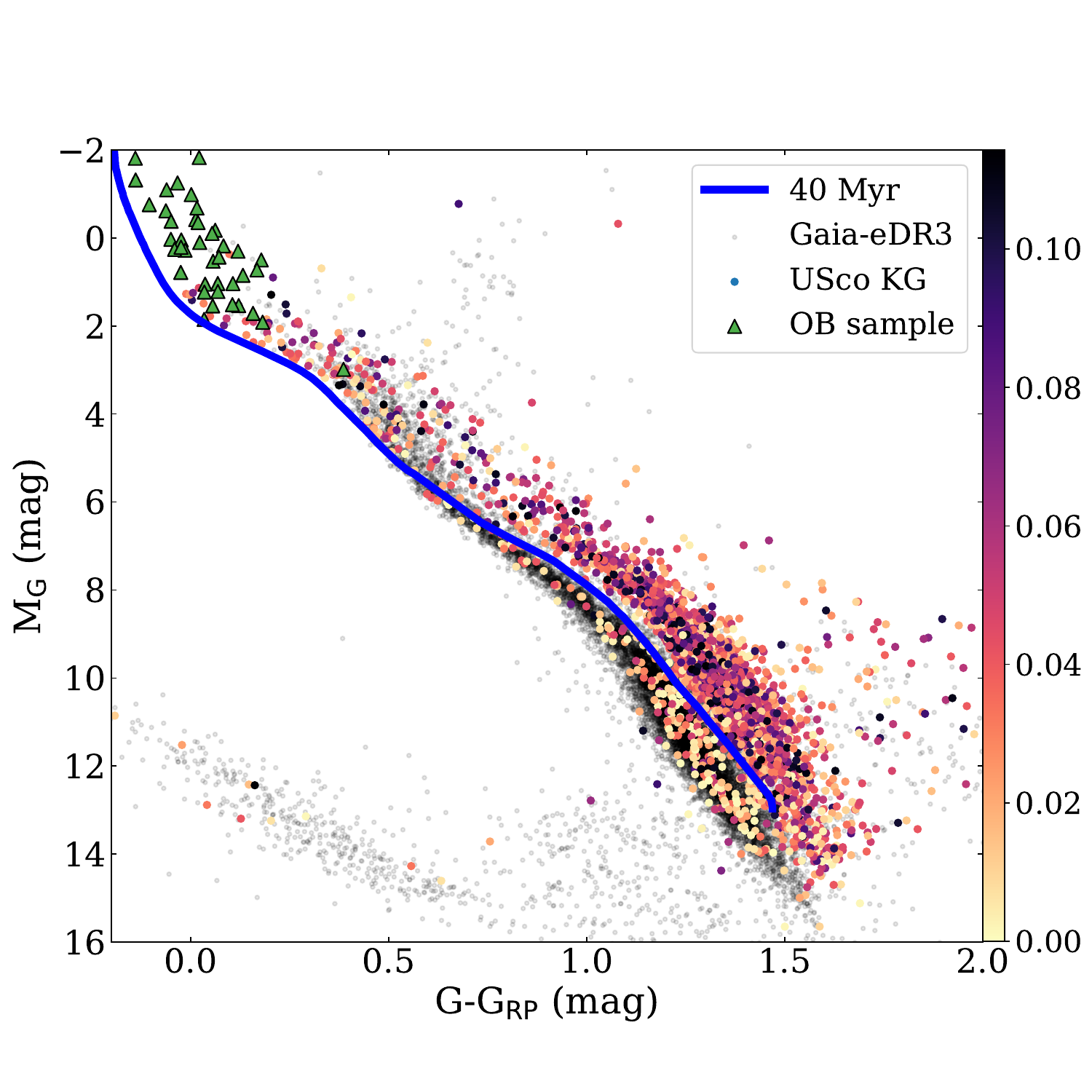}
    \caption{Color index G-G$_{\mathrm{RP}}$ vs absolute G magnitude for the Gaia-eDR3 set and the USco KG. The color map informs about the estimated posterior probability for each star to belong to the USco KG, the same as right panel in Figure \ref{figR.2}. A 40 Myr PARSEC isochrone divides the young low-mass pre-MS sources of USco from the MS stars of the Galactic field stars. The 47 stars with Gaia measurements from the OB sample are plotted as blue triangles. }
    \label{figR.4}
\end{figure}

\subsection{Main Kinematic Groups}\label{4.2} 

\begin{figure*}
    \centering
    \includegraphics[width=15cm]{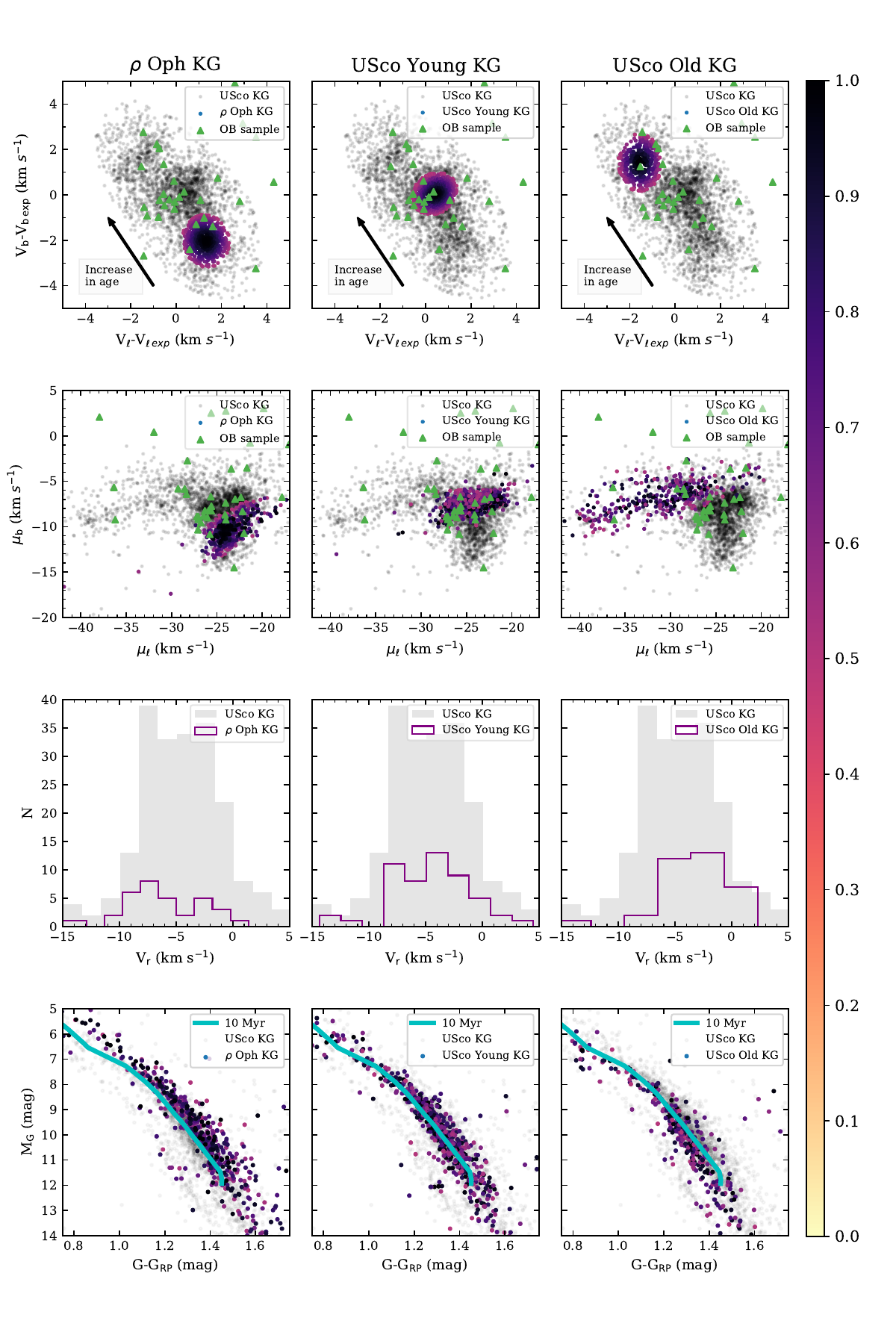}
    \caption{Columns: from left to right, the $\rho$ Oph, USco Young and USco Old KGs. \textbf{Rows:} 1st:  $\boldsymbol{\mathrm{V}}_\mathrm{T}*$ space. 2nd: proper motion space. 3rd: V$_\mathrm{r}$ distribution. 4th: CMD with a 10 Myr PARSEC isochrone. The color maps the min-max normalized $\mathrm{P(KG_i}|\mathrm{V}_\mathrm{T}*)>$. Only sources with $\mathrm{P(KG_i}|\mathrm{V}_\mathrm{T}*)>$50\% are plotted. The USco KG is shown as the gray background and stars from the OB sample are shown as green triangles. A direct comparison can be done between the different groups and the spaces of interest, revealing a direct relation between position in the kinematic spaces and age.}
    \label{figR.5}
\end{figure*}

\begin{figure*}
    \centering
    \includegraphics[width=15cm]{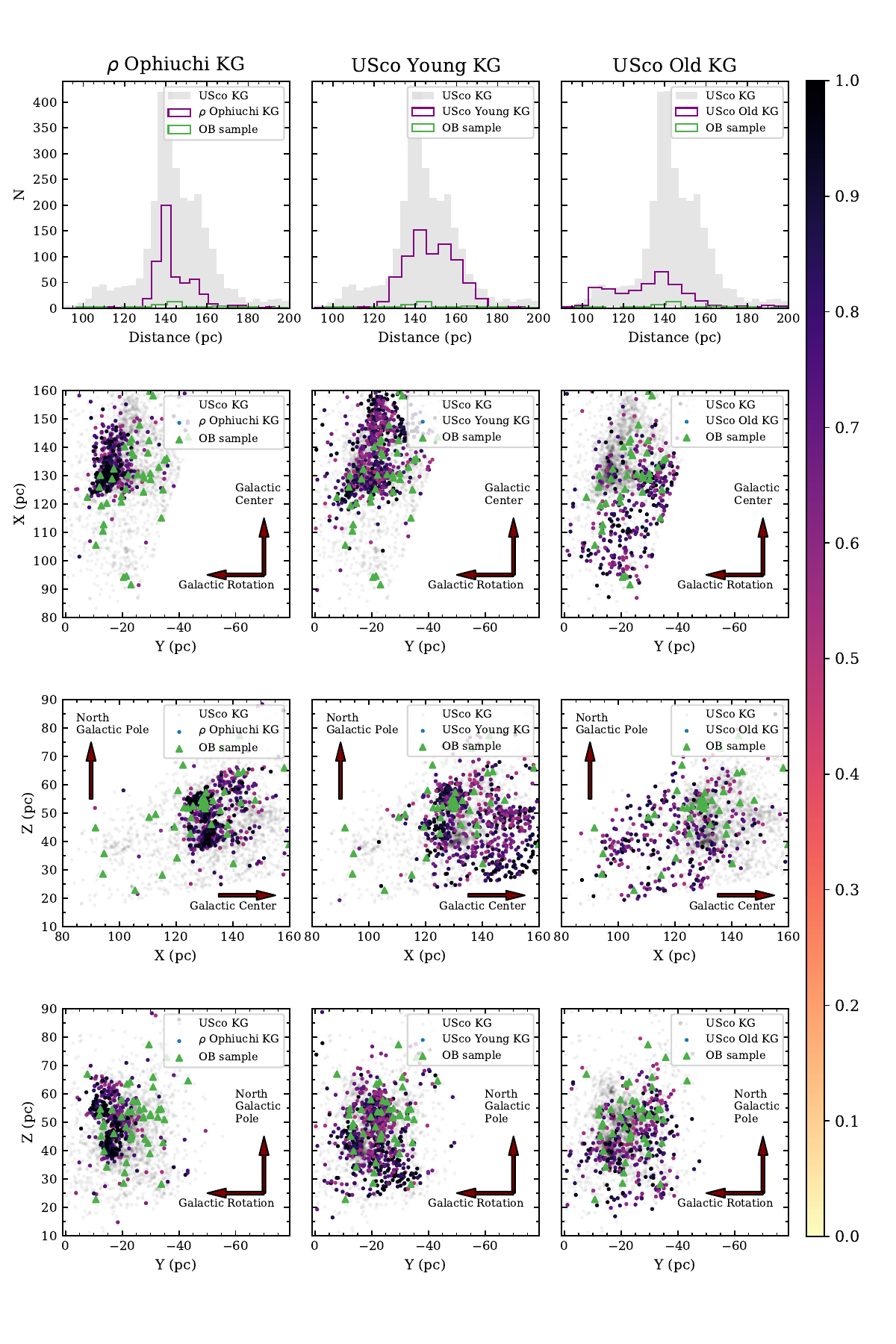}
    \caption{Columns: same as Figure \ref{figR.5}. Rows: 1st: Distance distribution. 2nd: Projection on the (Y,X) plane. 3rd: Projection on the (X,Z) plane. 4th: Projection on the (Y,Z) plane. Colors: same as Figure \ref{figR.5}. The distance distribution for the 3 KGs is double peaked, with all 3 groups showing a peak at $\sim$ 140 pc. Spatial distribution shows a complex structure for each KG, suggesting that clustering in kinematic spaces does not implies spatial clustering in USco and that spatial mixing between this KGs is non-negligible. The selected OB stars mostly overlap with the USco Young and USco Old KGs and tend to be in the external regions of USco.}
    \label{figR.6}
\end{figure*}

As very local over densities may have been missed by the global Second GM fit applied on the Gaia-eDR3 set (see Section \ref{3.1}), we perform a Third GM fit but on the USco KG (right panel in Figure \ref{figR.2}). The Third GM fit produces a minimized BIC if 3 components are chosen (see Figure \ref{AppendixC1}), suggesting that USco is actually composed by at least 3 kinematic substructures. We rule out that uncertainties significantly affect this result because typical propagation errors in V$_\alpha$ and V$_\delta$ are $\sim$ 0.1 and $\sim$ 0.2 km s$^{-1}$, one order of magnitude lower than the 1$\sigma$ regime of the narrowest $\mathrm{P(KG_i}|\mathrm{V}_\mathrm{T}*)$ distribution ($\sim$ 2.8 km s$^{-1}$) (see Section \ref{3.1.1}). In the following, we assess the physical relevance of the retrieved KGs and the limitations of this method. Figure \ref{figR.2}, summarizes the results from the Second and Third GM fits, i.e., left panel shows the Gaia-eDR3 set and the USco KG with the respective posterior probabilities (min-max normalized) extracted from the GM fits. We will refer to the 3 KGs retrieved as $\rho$ Oph KG, USco Young KG and USco Old KG. The $\rho$ Oph KG is mainly associated with the well known $\rho$ Ophiuchi region, the USco Young KG contains what typically has been taken as the remaining populations of USco and we present the USco Old KG as a new KG which remains unstudied as such. We will show that the inclusion of the USco Old KG is necessary to understand many important properties of USco. These KGs have mean velocities (U,V,W) $=$ ($-0.22\pm3.6, -16.2\pm0.7, -6.7\pm1.8$), ($-1.3\pm3.6, -17.0\pm0.9, -5.1\pm1.7$), ($-2.6\pm7.3, -18.8\pm0.9, -4.0\pm2.2$) km s$^{-1}$, for the $\rho$ Oph, USco Young and USco Old KGs, respectively.

In Figure \ref{figR.4} we show the G-G$_{\mathrm{RP}}$ vs M$_\mathrm{G}$ CMD for both the Gaia-eDR3 set and the USco KG. The color map indicates the posterior probability for each source to belong to the USco KG. A 40 Myr PARSEC \citep{parsecBressan2012,parseclowmasschen2014,parsecnewmarigo2017} isochrone is plotted to illustrate the division between the USco members from the Galactic field stars, from this, we note that low probability sources are also very likely to belong to the Galactic field populations, but also to the young stellar objects commonly associated with the $\rho$ Oph complex. Table \ref{Tabla1} show the estimates of the mixing percentiles for each KG, including USco (see Section \ref{3.1.1}). Rows show the USco KG and the remaining kinematic substructures in the confidence intervals selected, while columns indicate how many stars from a given KG are expected to be in the given confidence intervals for each row.

\begin{table}
    \centering
    \begin{tabular}{llll}
    \hline
    \hline
    Kinematic Group     & $\Sigma P(\rho|\mathrm{V}_\mathrm{T}*$)   & $\Sigma P(Young| \mathrm{V}_\mathrm{T}*)$   & $\Sigma P(Old| \mathrm{V}_\mathrm{T}*)$  \\
    \hline
    USco      &  35.6 $\%$          & 36.8 $\%$                 &    27.6 $\%$            \\
    $\rho$\:Oph    & --             & 8.2 $\%$ & 0.2 $\%$ \\  
    USco\:Young  & 14.9 $\%$            & -- & 9.1$\%$ \\
    USco\:Old      & 0.2 $\%$            & 4.4 $\%$& -- \\
    \hline
    \end{tabular}
    \caption{Rows: USco KG and the KGs retrieved with the Third GM fit with $\mathrm{P(KG_i}|\mathrm{V}_\mathrm{T}*)>50\%$ (min-max normalized). Columns: \%$_{mix}$ for each KG in the given regions.}
    \label{Tabla1}
\end{table}

The Figure \ref{figR.5} shows the sources with $\mathrm{P(KG}|\mathrm{V}_\mathrm{T}*) > 50\%$ for each KG\footnote{$\mathrm{P(KG_i}|\mathrm{V}_\mathrm{T}*)$ is min-max normalized.} described by the color map. 644, 702 and 467 sources are retrieved for the $\rho$ Oph, USco Young and USco Old KGs respectively. The USco KG is shown as gray background such that for each KG in USco (columns) a direct comparison on the parameters (rows) can be made. 
The first and second rows show the $\boldsymbol{\mathrm{V}_\mathrm{T}*}$ space, where the fit was made, and the observed proper motion, respectively. The third row shows the V$_\mathrm{r}$ histogram with the 36 ($\rho$ KG), 49 (USco Young KG) and 44 (USco Old KG) sources with available V$_\mathrm{r}$ measurements in the interval [-15,5] km s$^{-1}$, the means for the sources in the entire range is (--8.1, --3.9, --2.8 km s$^{-1}$) suggesting that the groups are also separated in this directly observed space, however, the typical error in V$_\mathrm{r}$ for the entire USco KG is $\sim$ 2.79 km s$^{-1}$ and we can not rule out field star contamination or the intrinsic internal mixing between groups as the cause of this separation. The last row shows the CMD for the absolute G magnitude (M$_\mathrm{G}$) and color index G-G$_{\mathrm{RP}}$ for the three KGs, each with a 10 Myr PARSEC isochrone for a relative age comparison between the groups. This CMDs suggest an age gradient between the groups. Note that the extinction vector is $\sim$ parallel to the isochrones in the color range shown, hence, extinction affected sources are displaced approximately along the isochrones implying that ages inferred from this CMD are confident enough to at least extract information of the relative ages of these stellar populations. The overall kinematic properties suggest a true physical distinction between the $\rho$ Oph, USco Young and USco Old KGs and from this CMDs we infer an unambiguous age distinction between the $\rho$ Oph and USco Old groups, this is consistent with the results from the kinematic spaces and serves as an independent confirmation of $\rho$ Oph KG and USco Old KG as distinct physical groups. Besides, supports the conclusion that the kinematic mixing between the USco Young KG and the other populations is non-negligible. Furthermore, the first row in Figure \ref{figR.6} shows a double peaked distance distribution for each KG. As the typical propagation error is $\sim$ 2 pc ($<$ 3 pc for $\sim$ 80$\%$ of the sources), the 2 peaks for the KGs are likely to be real. The rows bellow, which are projections in the (Y,X), (X,Z) and (Y,Z) Galactic sub-spaces, help to clarify this; for the $\rho$ Oph KG, at least 3 rather compact substructures are revealed (X,Y,Z $\sim$ (132,-18,42), (128,-12,55), (140,-17,60) pc), mainly, in the (X,Z) and (Y,Z) projections, 2 of them explain the distance peak at 140 pc (X $\sim$ 130 pc). In the case of the USco Young KG, 2 rather definite substructures appear at (X,Y,Z $\sim$ (130,-22,55),(128,-15,43) pc) and a third diffuse covering the region (X $>$ 135 pc, Y $<$ -19 pc, Z $<$ 52 pc). Finally, the USco Old KG is mainly constituted by 1 small apparent substructure (X,Y,Z $\sim$ (100, -20, 38) pc) and a widely spread Diffuse population towards the east of USco, connecting it with the Upper Centaurus Lupus (UCL) region (see \cite{Damiani2019}, its Figure 37).

Analogue procedures to the Kinematic Analysis, have been used as the main tool in recent works \citeg{Luhman2018,Luhman2020,Luhman2022} to investigate the kinematic substructure. However, this result suggests that the GM fit does not completely disentangles the spatial or age substructure of USco, i.e., there may be true substructures with too similar kinematics as to be separated via this method. 

On the other hand, the USco Old KG has been left out of some recent studies on USco based on the Gaia data, for example, the previous works by \cite{Luhman2020}, \cite{Damiani2019} and \cite{kerr} with Gaia-DR2 data have omitted the inclusion of the USco Old KG mainly due to: 1. They apply clustering techniques like GM but in the proper motion offset space, where uncertainties are lower but, due to its large distance dispersion and lower density, the USco Old KG does not appear clustered, so that on a conservative confidence intervals it may be classified as contamination. This was indeed the case for the analysis of \cite{Luhman2020} (see its Figure 6). 2. In the work of \cite{Damiani2019}, the selection method produces a rather complete final sample and contains most sources from the USco Old KG, but no systematic way of look for KGs in the V$_\mathrm{T}$ space in USco is performed (see its Figs.12,13,14). 3. The clustering space investigated with HDBSCAN by \cite{kerr} requires both spatial and kinematic clustering, allowing them to recover the substructure SC-13 but as completely separated from USco, when it is probably related to the SC-17-A substructure. 

\begin{table}
    \caption{Candidate members of the Upper Scorpius Kinematic Group.} \label{Table1.5}
    \begin{threeparttable}
        \begin{tabular}{l|l}
            \hline
            \hline
            Column Label   & Description  \\
            \hline 
            source\_id                      & \\
            ra                              & Right ascension from Gaia eDR3 \\
            ra\_error                       & Error in ra from Gaia eDR3 \\
            dec                             & Declination from from Gaia eDR3 \\
            dec\_error                      & Error in dec from Gaia eDR3 \\
            parallax                        & Parallax from Gaia eDR3 \\
            parallax\_error                 & Error in parallax from Gaia eDR3 \\
            distance                        & Distance by inversion of parallax \\
            pmra                            & Proper motion in ra from Gaia eDR3\\
            pmra\_error                     & Error in pmra from Gaia eDR3\\
            pmdec                           & Proper motion in dec from Gaia eDR3 \\
            pmdec\_error                    & Error in pmdec from Gaia eDR3\\
            ipd\_gof\_harmonic\_amplitude   & Amplitude of the Image Parameter \\ 
                                            & Determination Goodness of Fit \\ 
                                            & versus position angle of scan \\
            ipd\_frac\_multi\_peak          & Percent of successful-Image Parameter \\ 
                                            & Determination windows with more than \\ 
                                            & one peak \\
            ruwe                            & Renormalized unit weight error \\ 
                                            & from Gaia eDR3 \\
            phot\_g\_mean\_mag              & Magnitude in the G band \\ 
                                            & from Gaia eDR3 \\
            phot\_rp\_mean\_mag             & Magnitude in the G$_{RP}$ band \\ 
                                            & from Gaia eDR3 \\
            M\_g                            & Absolute magnitude in the G band \\
            dr2\_radial\_velocity           & Radial velocity from Gaia DR2 \\
            dr2\_radial\_velocity\_error    & Error in radial velocity \\ 
                                            & from Gaia DR2 \\
            l                               & Galactic longitude \\
            b                               & Galactic latitude \\
            pml                             & Proper motion in the $\ell$ direction \\
            pmb                             & Proper motion in the $b$ direction \\
            vl                              & Tangential velocity in the $\ell$ \\
                                            & direction \\
            vb                              & Tangential velocity in the $b$ \\
                                            & coordinate \\
            vl\_exp                         & Expected tangential velocity \\ 
                                            & in the $\ell$ direction \\
            vb\_exp                         & Expected tangential velocity \\ 
                                            & in the $b$ direction \\
            X                               & Cartesian heliocentric coordinate \\ 
                                            & in the direction of the Galactic \\ 
                                            & center \\
            Y                               & Cartesian heliocentric coordinate \\ 
                                            & in the direction of the Galactic \\ 
                                            & rotation \\
            Z                               & Cartesian heliocentric coordinate \\ 
                                            & in the direction of the North \\ 
                                            & Galactic Pole \\
            p\_rho\_Oph\_kg                 & $\rho$ Ophiuchi KG posterior proba- \\
                                            & bility from the third GM fit \\
            p\_usco\_young\_kg              & USco Young KG posterior proba- \\ 
                                            & bility from the third GM fit \\
            p\_usco\_old\_kg                & USco Old KG posterior proba- \\ 
                                            & bility from the third GM fit \\
            epsilon                         & $\epsilon$ value yielded by OPTICS \\
            V\_core                         & V$_{core}$ from Equation \ref{eqVcore}  \\
            substructure                    & Label assigned by OPTICS  \\
            \hline
        \end{tabular}
        \begin{tablenotes}
          \small
          \item Labels assigned by OPTICS: -1: Diffuse populations, 0: $\pi$ Sco, 1: $\rho$ Oph, 2: $\omega$ Sco, 3: $\nu$ Sco, 4: $\delta$ Sco, 5: $\alpha$ Sco, 6: $\beta$ Sco, 7: UCL\'s. Data from this table will be provided as supplementary material to this paper.
        \end{tablenotes}
    \end{threeparttable}
\end{table}

\begin{figure*}
    \centering
    \includegraphics[width=17cm]{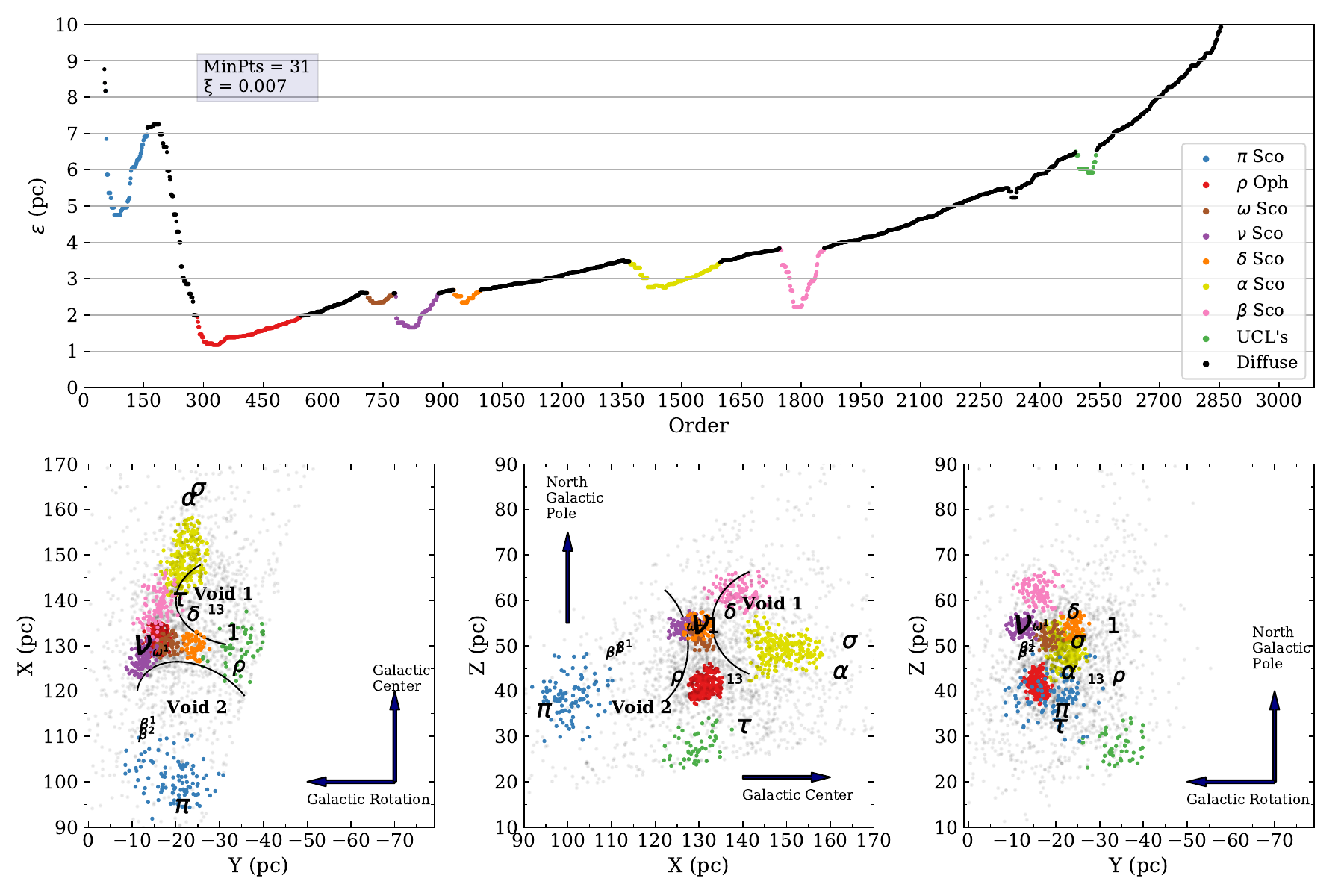}
    \caption{Top: R plot for the USco KG with $MinPts=$31 and $\xi=$0.07). Colored valleys represent each structure classified as a cluster in (X.Y,Z) space by the OPTICS algorithm with the given parameters. Diffuse populations are identified as black plateaus. Bottom: from left to right, projections on the Galactic (Y,X), (X,Z) and (Y,Z) planes. Colors represent each substructure extracted from the R plot. The Diffuse populations are plotted in gray. The stellar voids observed are shown as elliptical arcs where such geometry can be clearly assigned. The position of stars from the Turn-off sample is represented by the respective Bayer designation. Note that the Turn-off sample does not include the star $\rho$ Ophiuchi, hence, the $\rho$ symbol in this and the following figures corresponds to the star $\rho$ Scorpii. }
    \label{figR.7}
\end{figure*}

\begin{table*} 
    \begin{tabular}{lllllllll}
    \hline
    \hline
    Substructure                & $\rho$ Oph      & $\alpha$ Sco  & $\beta$ Sco   & $\pi$ Sco     & $\nu$ Sco      & $\omega$ Sco  & $\delta$ Sco  & Diffuse \\
    \hline 
    N$_\star$                   & 261             & 227           & 111           & 104           & 108            & 66            & 66            &       2009          \\
    $N_{V_\mathrm{r}}$          & 16              & 14            & 4             & 15            & 7              & 4             & 7             & 188 \\
    $N_{A_G}$                   & 47              & 165           & 106           & 20            & 103            & 59            & 59            & 835 \\
    $\alpha$ [deg]              & 246.4$\pm0.6$   & 244.5$\pm1.0$ & 241.3$\pm0.8$ & 240.7$\pm2.7$ & 242.7$\pm0.5$  & 242.1$\pm0.8$ & 239.8$\pm0.7$ & 243.6$\pm4.6$ \\
    $\delta$ [deg]              & -24.1$\pm0.7$   & -24.6$\pm1.1$ & -19.5$\pm0.9$ & -24.5$\pm2.3$ & -19.4$\pm0.6$  & -21.9$\pm0.7$ & -23.1$\pm0.6$ & -24.6$\pm3.8$ \\
    $\varphi$ [mas]                 & 7.2$\pm0.1$     & 6.3$\pm0.2$   & 6.5$\pm0.1$   & 9.1$\pm0.3$   & 7.2$\pm0.1$    & 7.1$\pm0.05$  & 7.0$\pm0.1$   & 7.2$\pm1.8$   \\
    $d$ [pc]                    & 138.7$\pm2.0$   & 158.2$\pm4.1$ & 152.8$\pm3.0$ & 109.9$\pm4.3$ & 138.8$\pm2.1$  & 141.4$\pm1.1$ & 143.0$\pm1.7$ & 144.2$\pm21.4$\\
    $X$ [pc]                    & 131.4$\pm1.8$   & 148.7$\pm4.3$ & 138.8$\pm3.2$ & 101.0$\pm4.3$ & 127.2$\pm2.0$  & 130.1$\pm1.4$ & 130.0$\pm1.7$ & 134.2$\pm20.3$\\
    $Y$ [pc]                    & -15.8$\pm1.2$   & -22.0$\pm2.3$ & -16.0$\pm2.1$ & -19.4$\pm5.0$ & -12.4$\pm1.5$  & -18.2$\pm1.5$ & -24.1$\pm1.3$ & -21.7$\pm9.3$ \\
    $Z$ [pc]                    & 41.1$\pm1.8$    & 49.4$\pm2.8$  & 61.7$\pm2.0$  & 38.8$\pm4.1$  & 54.2$\pm1.3$   & 52.2$\pm1.7$  & 54.3$\pm1.6$  & 46.2$\pm12.8$ \\
    $\mu_{\alpha}$ [mas yr$^-1$]& -7.6$\pm2.0$    & -10.3$\pm1.2$ & -9.8$\pm1.1$  & -20.3$\pm2.5$ & -9.2$\pm1.4$   & -10.4$\pm1.0$ & -12.6$\pm1.0$ & -12.2$\pm4.8$ \\
    $\mu_{\delta}$ [mas yr$^-1$]& -25.7$\pm1.5$   & -21.8$\pm1.3$ & -21.7$\pm0.7$ & -31.8$\pm2.4$ & -24.0$\pm1.1$  & -24.1$\pm1.0$ & -24.0$\pm0.9$ & -24.5$\pm6.5$ \\
    v$_\mathrm{r}$ [km s$^-1$]  & -0.42$\pm1.8$   & 4.8$\pm18.9$  & -8.8$\pm4.6$  & -0.2$\pm2.7$  & -8.0$\pm1.2$   & -1.9$\pm7.6$  & -7.0$\pm1,2$  & -6.5$\pm17.8$ \\
    $U$ [km s$^-1$]             & -0.1$\pm1.7$    & 3.9$\pm17.6$  & -7.8$\pm4.3$  & -2.4$\pm2.6$  & -6.6$\pm1.8$   & -1.3$\pm7.1$  & -7.5$\pm1.0$  & -7.4$\pm16.6$ \\
    $V$ [km s$^-1$]             & -16.0$\pm0.9$   & -18.0$\pm3.3$ & -15.8$\pm0.9$ & -19.0$\pm1.1$ & -15.2$\pm0.8$  & -16.3$\pm1.2$ & -16.4$\pm0.3$ & -16.6$\pm2.6$ \\
    $W$ [km s$^-1$]             & -7.2$\pm1.5$    & -3.9$\pm5.8$  & -8.7$\pm1.7$  & -4.1$\pm1.5$  & -8.4$\pm0.7$   & -6.8$\pm2.2$  & -7.8$\pm0.6$  & -7.0$\pm6.0$  \\
    $A_G$ [mag]                 & 1.7$\pm0.76$    & 1.2$\pm0.94$  & 0.51$\pm0.30$ & 0.20$\pm0.20$ & 0.78$\pm0.58$  & 0.55$\pm0.29$ & 0.44$\pm0.34$ & 0.58$\pm0.62$ \\

    \hline
    \end{tabular}
    \caption{Mean astrometric parameters and extinction in the G band, with 1$\sigma$ standard deviations for the seven substructures of USco and the Diffuse population identified with OPTICS. The number of available sources with radial velocity and extinction values is also listed.} 
    \label{Table2}
\end{table*}

\begin{figure*}
    \centering
    \includegraphics[width=\textwidth]{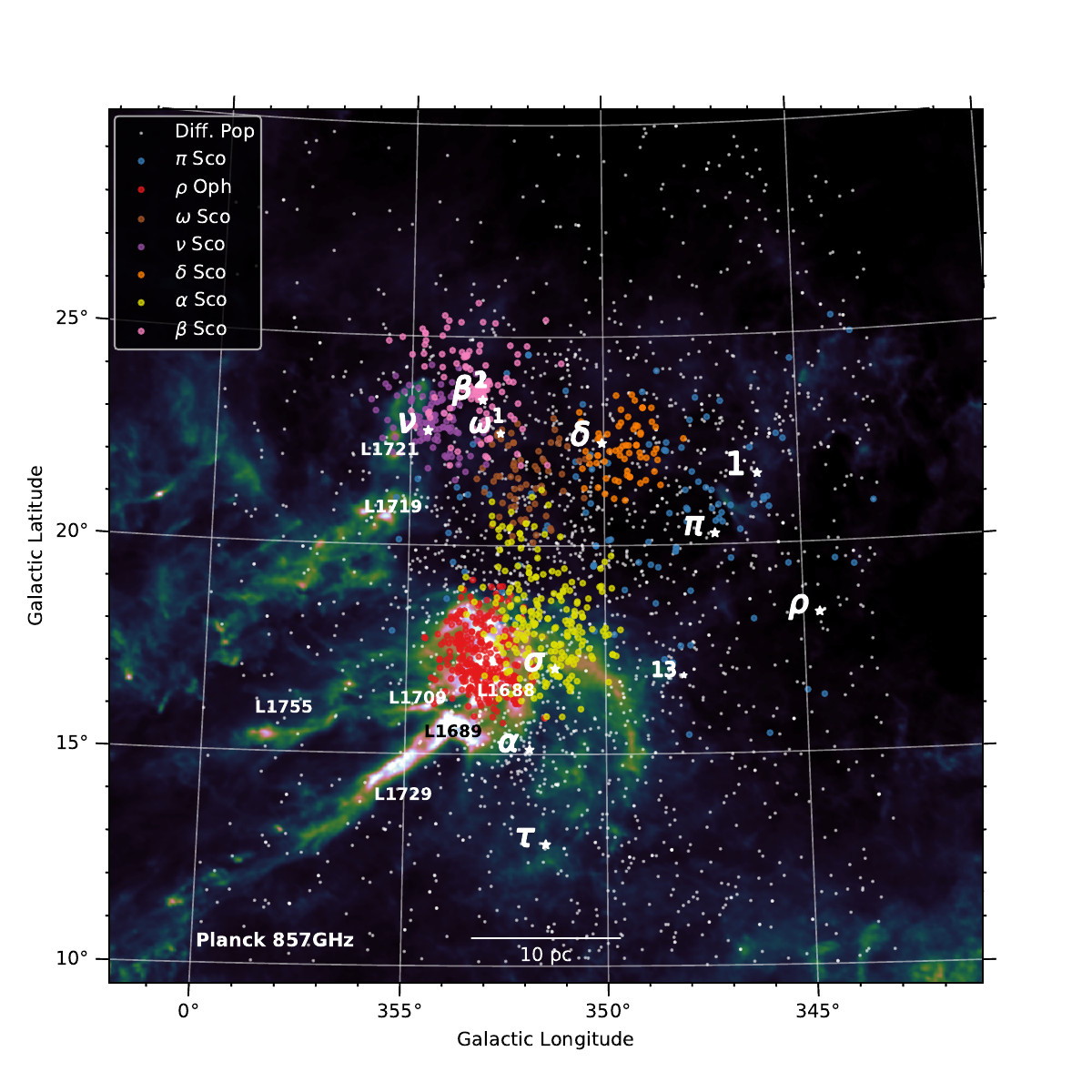}
    \caption{Celestial positions in galactic ($\ell$,b) coordinates for each clustered substructure (colors) and the Diffuse populations (light gray) retrieved with OPTICS. The stars of the Turn-off sample are plotted as with star symbols with their Bayer designation at the left. The background is an image from Planck at 857 GHz in the USco region and allows to see the dust emission related to HI structures and molecular clouds. The main Lynds clouds are identified. The scale of 10 pc is at d $\sim$ 140 pc.} 
    \label{figR.8}
\end{figure*}

\begin{figure*}
    \centering
    \includegraphics[width=\textwidth]{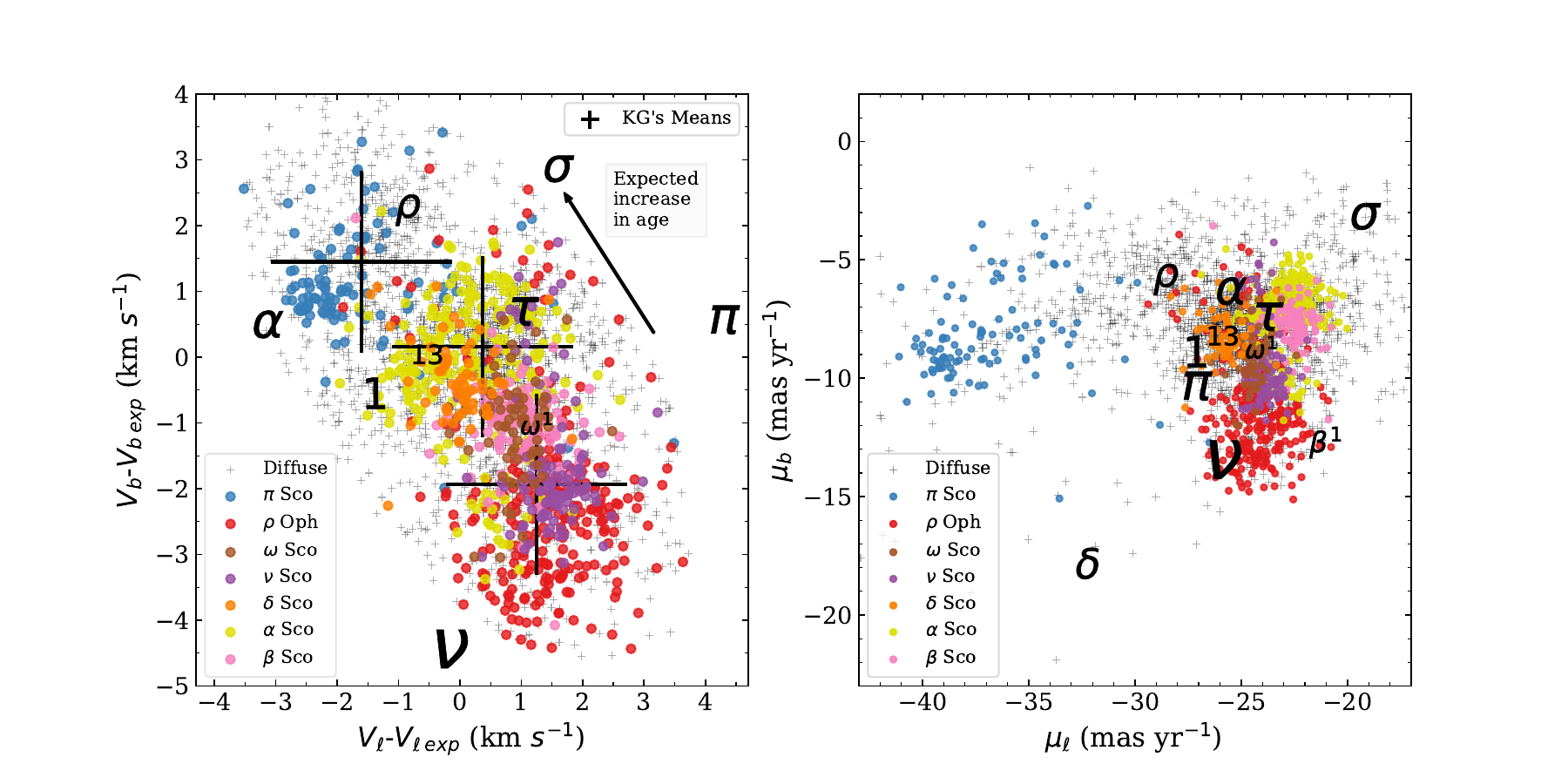}
    \caption{Left: $\boldsymbol{\mathrm{V}}_{\mathrm{T}}*$ space. The 7 substructures retrieved with OPTICS are shown. We plot the expected direction where the age increases and crosses centered on the mean positions of the $\rho$, $\alpha$ and $\pi$ KGs, according to the results from the Kinematic Analysis. The crosses have width $\sim$ 1$\sigma$ of the fitted components. Right: proper motion. Notably, these spatially clustered populations are also highly clustered in these kinematic spaces. Correcting perspective and depth effects is important, as the trend and relative positions of the substructures are better extracted from the $\boldsymbol{\mathrm{V}}_{\mathrm{T}}*$ space. }
    \label{figR.9}
\end{figure*}

\subsection{Internal structure of USco}\label{4.3} 

In following sections we show how our OPTICS implementation in the Gaia eDR3 data verifies and significantly complements the previous investigation of the substructure in USco by producing new and more robust results about it.

\subsubsection{Spatial distribution}\label{4.3.1}

 The parameter selection of the OPTICS algorithm largely depends on the intended resolution of the analysis to perform. We limit our analysis to identify spatial substructures in the USco KG, if present, containing at least 31 sources ($\sim 1\%$ of our USco KG). We explored the ($MinPts$, $\xi$) parameters in the ranges [2,41], [0.006,0.06] and report the results of choosing [$MinPts = 31$, $\xi = 0.007$]. Varying $MinPts$ from 21 to 41 does not produce any significant change in the global or local shape of the R plot (see Section \ref{3.2.2}), values below 21 produce noisy steepness in the already present valleys, which makes difficult to choose $\xi$ when assigning cluster limits, but the global structure is roughly the same. Values below 21 start to introduce too local valleys with typically $\sim$ 10 sources introducing noise and smaller substructures. The structure most affected by this changes is the labeled as UCL's for its spatial position, as it disappears and strongly changes its shape when varying this parameter, hence, we omit its inclusion as a well defined structure and also any further analysis of it. As $MinPts$ only needs to be large enough to distinguish the substructures with the intended resolution \citep{OPTICS}, we chose $MinPts = 31$ and discarded the analysis of $MinPts > 41$. Once the $MinPts$ is set, we start to vary $\xi$ from 0 to the maximum value where all the valleys with at least 31 sources are selected as clusters ($\xi = 0.005$) in $0.001$ steps. Our main analysis is based on $\xi = 0.007$ as it is the first non-trivial value for $\xi$, given $MinPts$, we also study the effects of varying $\xi$ from 0.007 until the point where none of the valleys are selected as a substructure ($\xi = 0.06$). 

The R plot (see Section \ref{3.2.2}) produced by OPTICS is shown in the top panel of Figure \ref{figR.7}. Each source has a position in the R plot and each substructure satisfying the conditions above mentioned is shown in a different color, while the Diffuse populations in USco are identified with black dots. In Figure \ref{figR.7} we show the direct imprint of the spatial structure of USco and evidence for the presence of 8 substructures plus the Diffuse populations, where a total of 1049 and 2065 sources belong to the clustered and Diffuse populations, respectively. We suggest to name six of these structures after their brightest star (see footnote 1): $\pi$ Scorpii, $\omega$ Scorpii, $\nu$ Scorpii, $\delta$ Scorpii, $\alpha$ Scorpii and $\beta$ Scorpii.The relation between these structures and the respective stars is first shown in this work, and will be discussed in section \ref{4.5}. We also recover the well known $\rho$ Ophiuchi substructure and another minor substructure that we will label UCL's. 
From the work of \cite{kerr}, we recover the substructures SC-13 ($\pi$ Scorpii), SC-17-E,G,I ($\nu$,$\beta$ Scorpii and $\rho$ Ophiuchi) and increase their membership in at least 2 and up to 4 times the previous estimates. We find that the substructure H is actually composed by 2 substructures ($\omega$, $\delta$ Scorpii). SC-17-B,C,D are largely contained in $\alpha$ Scorpii substructure. We do not found evidence for the presence of the SC-F substructure and we argue that SC-A is more likely to be a spurious detection and part of the larger $\pi$ KG. Hence, we introduce 3 new substructures ($\omega$, $\delta$, $\alpha$ Scorpii) plus the physical relation of all the substructures with one of the most massive members in USco.
Table \ref{Table2} shows the mean values of the main astrometric variables for each of the 8 analyzed substructures\footnote{We omit the analysis of the UCL\'s substructure (51 sources) because of its noisy character and its truncated position towards the Upper Centaurus Lupus region.} and their respective 1$\sigma$ standard deviations. The lower panels in Figure \ref{figR.7} show the spatial projections of the five major substructures with $> 100$ sources, the other two with 66 sources each and the Diffuse populations (gray background) in the Galactic (X,Y,Z) coordinate sub-spaces plus the stars in the Turn-Off sample with white star symbols. Figure \ref{figR.7} show unprecedented insights on the internal structure of USco. Among the most notable characteristics from the spatial distribution of this substructures are the apparent voids shown in the lower panels of Figure \ref{figR.7}. These voids are apparently well fitted by ellipsoids connected at their edges (the arcs shown in Y-X and X-Z projections) by the $\rho$ Oph, $\beta$ Sco, $\omega$ Sco, $\nu$ Sco, and $\delta$ Sco substructures. We will revisit the relation between the Voids and the substructures when analyzing the star formation history of USco in Section \ref{5.3}.

To obtain a quantitative idea of the reliability of these results, we base on the OPTICS' $\xi$ parameter. When varying the $\xi$ parameter, we note that each substructure begins to lose members when the gradient of the valley in the R plot is too low with respect to to $\xi$, this means that changing the $\xi$ parameter can give us an idea on how definite are the boundaries of a cluster relative to its surroundings. Paradoxically, the first substructure in losing members, at $\xi$ $=$ 0.007, is the associated with $\rho$ Oph (red valley in Figure \ref{figR.7}), the most deeply studied region in USco. This by itself is evidence of the true physical relevance of the new substructures extracted with OPTICS as they are more robustly clustered than $\rho$ Ophiuchi. On the other hand, we note by the same methods that the $\beta$ Sco and $\pi$ Sco substructures are the most well defined substructures retrieved, which is also noted from their higher steepness in the R plot. 
The celestial ($\ell$,$\mathrm{b}$) position of these substructures and the Diffuse populations is shown in Figure \ref{figR.8}. A Planck 857 GHz dust emission background is also plotted with visualization purposes; the dust emission maps very well the HI structures at low densities and shows an excess of about 40\% with respect to to the HI emission in high density regions where molecular clouds are also present \citep{plackdust}, several Lynds clouds are also shown with this purpose. We also mark the positions of the brightest members of USco.

\subsubsection{Kinematics}\label{4.3.2}

The kinematic properties of this substructures can be seen in Figure \ref{figR.9}, where we show the 7 substructures in the $\boldsymbol{\mathrm{V}_\mathrm{T}*}$ (left) and proper motion (right) spaces. We include the Turn-Off sample in this plots, however, this proper motion and parallax measurements have non-negligible errors because they are strongly affected by binary interactions, deviating their kinematics from that of their parent substructures. These shows how the inclusion of transverse velocities would bias and reduce our member selection due to the unresolved binary interactions. Note that several substructures do not overlap with each other within the proper motion space as clearly as they do in the $\boldsymbol{\mathrm{V}_\mathrm{T}*}$ space. In particular, these effects are needed to be taken into consideration for the investigation of the Wall of USco. We also include the direction towards we expect for the age of all the substructures to increase, based on results from section \ref{4.2}. In contrast with the results obtained in section \ref{4.2}, substructures are compact and well defined in both (X,Y,Z) and $\boldsymbol{\mathrm{V}_\mathrm{T}*}$ spaces. Moreover, substructures are aligned in roughly the same direction the Kinematic Analysis suggests an increase in age with respect to the $\boldsymbol{\mathrm{V}_\mathrm{T}*}$ space. The 7 substructures constitute four clearly evident sub KGs, where four of them unambiguously overlaps with the high confidence region of the three KGs obtained in the Kinematic Analysis and the remaining two, $\beta$ and $\omega$ Sco, are located between two of these KGs, providing a way of testing how intrinsic is the age-$\boldsymbol{\mathrm{V}_\mathrm{T}*}$ dependency suggested by the Kinematic Analysis. We also note that $\delta$, $\beta$, $\omega$ and $\nu$ Sco are particularly close in both (X,Y,Z) and $\boldsymbol{\mathrm{V}_\mathrm{T}*}$ spaces a present a correlated directionality in these spaces. In the following, we try to obtain some insights on the age trend observed in the Kinematic Analysis in terms of substructures discussed.

\begin{figure*}
    \centering
    \includegraphics[width=18cm]{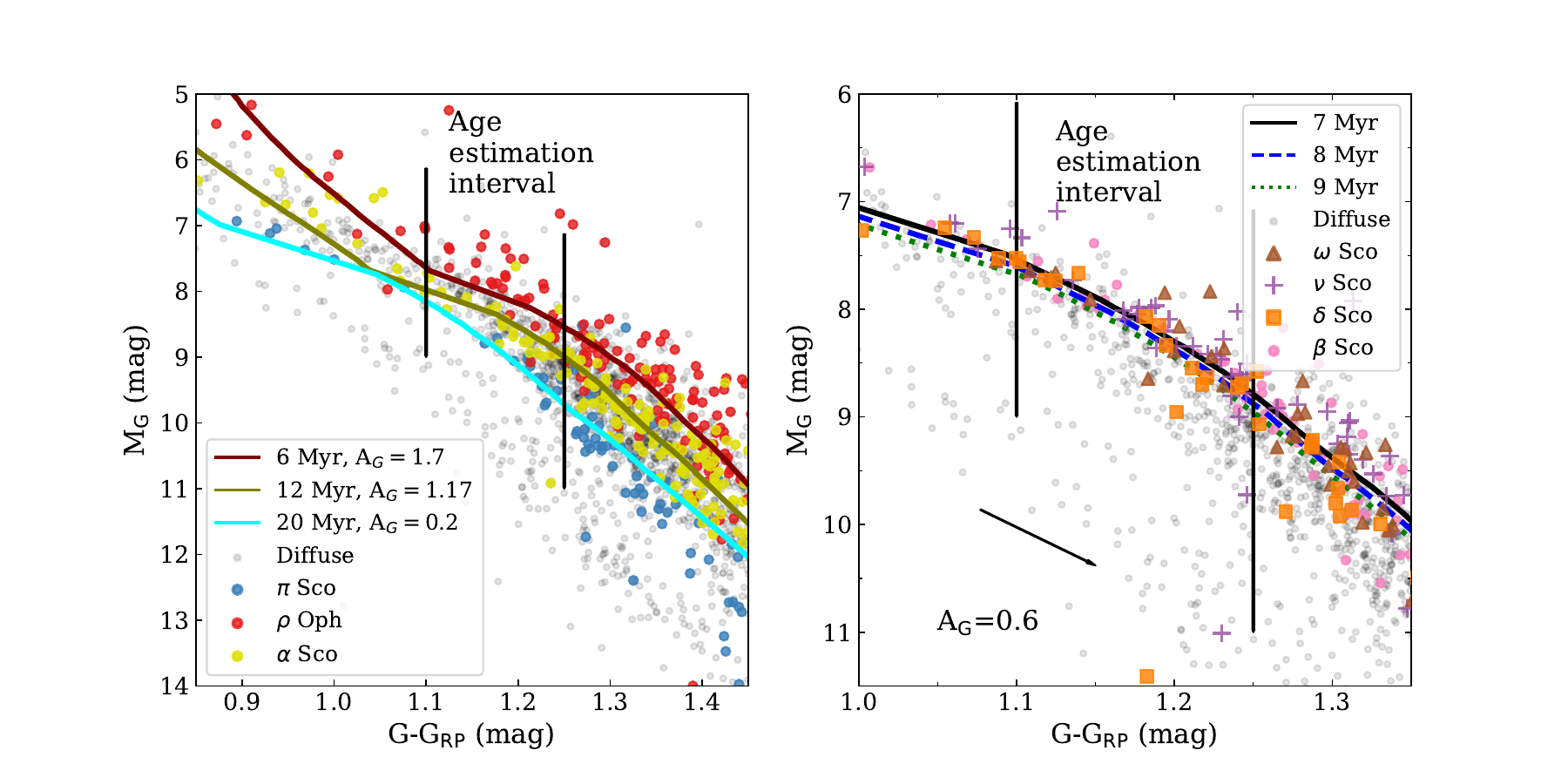}
    \caption{Left: CMD for the $\rho$, $\alpha$, $\pi$ Sco substructures and the Diffuse populations with G-G$_{\mathrm{RP}}$ colors and absolute G magnitude M$_\mathrm{G}$. The maximum age of $\rho$ Oph and the mean age of $\alpha$ Sco and $\pi$ Sco are fitted by PARSEC extinction-corrected 6, 14 and 20 Myr isochrones. Right: same as left but for the $\nu$, $\beta$, $\omega$ and $\delta$ Sco substructures. Mean ages of these substructures are fitted with PARSEC extinction-corrected (A$_\mathrm{G}$=0.6) 7, 8, 8 and 9 Myr isochrones. We include extinction vectors to illustrate the average effects of extinction. In both panels, vertical bars at G-G$_{\mathrm{RP}}=$1.1 and G-G$_{\mathrm{RP}}=$1.25 delimit our age estimation interval (see text for details). } \label{figR.10}
\end{figure*}

\subsubsection{Ages}\label{4.3.3}

Here we make relative and intrinsic estimations for the substructures found in section \ref{4.3}. In order to do this, we use PARSEC \citep{parsecBressan2012,parseclowmasschen2014,parsecnewmarigo2017} isochrones and the G, G$_{\mathrm{RP}}$ Gaia-eDR3 passbands, as they exhibit lower extinction and errors than G$_{\mathrm{BP}}$. We aim to provide an insight into the age properties of this substructures to see if the expected age trend shown in Figure \ref{figR.9} is true for them, which may be tested extracting, if possible, relative ages between them from the CMD. While substructures show clearly distinct relative ages, in most cases, intrinsic age estimations are systematically affected, at least, by extinction effects at the individual level, hence, we include conservative systematic and statistic error estimations for the ages reported. \cite{Luhman2020} gives A$_\mathrm{K}$ (Ks passband from 2MASS) estimates for USco sources excluding the $\rho$ Oph region. Assuming a Cardelli reddening law \citep{extinctioncardelli1989}, we transform A$_\mathrm{K}$ values from \cite{Luhman2020} into A$_\mathrm{G}$. In Figure \ref{figR.10} we show the resulting CMDs with 6, 7, 8, 9, 14 and 20 Myr extinction-corrected isochrones. We assume a solar metallicity of Z$=$0.014 to be consistent with the grids from \cite{ekstromgrids} in the Time Analysis. A$_G$ mean values for each substructure are shown in Table \ref{Table2}. We take A$_G=$0.6 when correcting isochrones for $\delta$ Sco, $beta$ Sco, $\omega$ Sco and $\nu$ Sco since extinction does not vary significantly from this mean value for these substructues. 
To estimate intrinsic ages we base on a maximum age of $\rho$ Oph complex of 5-6 Myr \citep{Canovas2019,esplin2020}. The uncertainty in this value systematically propagates to our results. We chose a particular interval to estimate ages as different intervals yield different results. To chose the optimal interval we consider the following: 1. We aim to reproduce the 6 Myr upper age of $\rho$ Oph substructure. 2. Age estimations based on the region G-G$_{\mathrm{RP}}$ $<$ 1 would be strongly affected by low number statistics and unresolved binaries. 3. Current stellar evolutionary models are considered unreliable for late M objects \citep{PecautSco-cen2016}, in particular, observations suggest an underestimation of their radius \citeg{underestimateradius2012}, so we want to avoid as many late M stars as possible. By characterizing and constraining the fundamental properties of M stars, \cite{edades03508} concluded that M objects with mass $>$ 0.35 $M_\odot$ should be well described by current models at the time of Version 1.1 of the PARSEC grids, that we use in our work. Consistently, dispersion, deviation and the deviation gradient significantly increase at G-G$_{\mathrm{RP}}$ $>$ 1.25. At this mass regime, pre-MS M sources have an underestimated temperature by $\sim$ 100 K (see \cite{Pecaut2013SpT}, tables 5 and 6), hence, we estimate ages based on the G-G$_{\mathrm{RP}}$ interval [1.1,1.25], marked with vertical lines in both panels of Figure \ref{figR.10}. This interval would correspond to a mass range of $\sim$ [0.27-0.44] M$_\odot$ \citep{Pecaut2013SpT,Luhman2022}. We make age estimations for the remaining substructures relative to these considerations. 
From left panel in Figure \ref{figR.10} we see that the age trend observed in the Kinematic Analysis (see Figure \ref{figR.5}) holds for these substructures, i.e., $\rho$ Oph (6 Myr max), $\alpha$ Sco (14 Myr) and $\pi$ Sco (20 Myr), where all of this structures unambiguously belong to one of the KG found in the Kinematic Analysis. In the right panel of Figure \ref{figR.10}, we show the CMD for the $\nu$, $\omega$, $\beta$ and $\delta$ Sco substructures, as expected from their (X,Y,Z) and $\boldsymbol{\mathrm{V}_\mathrm{T}*}$ positions, we observe similar relative ages. We show 7, 8, 8 and 9 Myr isochrones for $\nu$, $\omega$, $\delta$ and $\delta$ Sco, respectively. We note that the global trend, better observed at the right of the age estimation interval, in the entire G-G$_{\mathrm{RP}}$ range is the expected from left panel of Figure \ref{figR.9}. 

\subsubsection{Age-velocity correlation}\label{4.3.4}

\begin{figure}
    \centering
    \includegraphics[width=8cm]{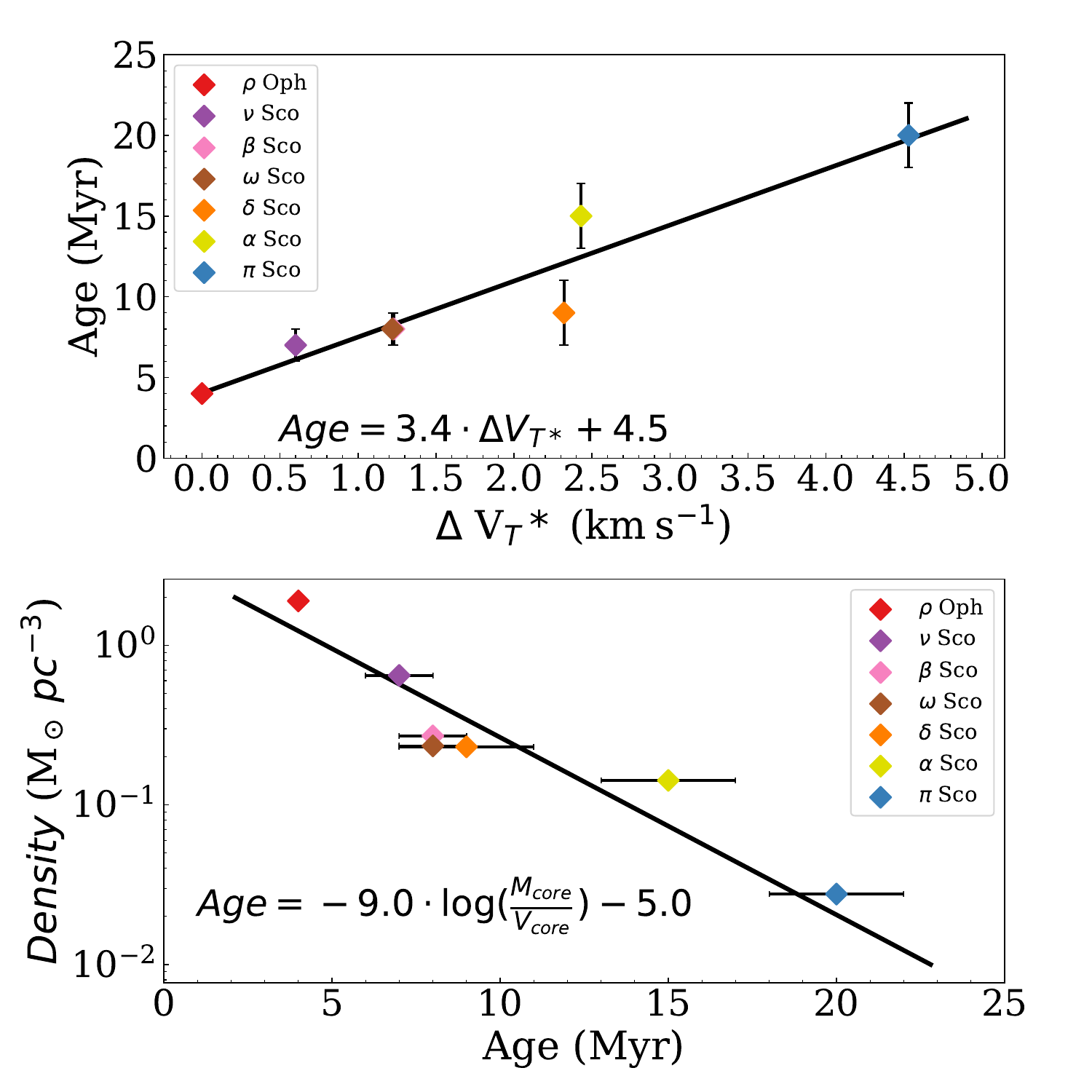}
    \caption{Top: $\Delta \boldsymbol{\mathrm{V}}_{\mathrm{T}}*$ vs age. Bottom:  age vs density. Upper and lower bounds for the photometric age estimates are shown.} \label{figR.11}
\end{figure}

To further investigate the intrinsic $\boldsymbol{\mathrm{V}_\mathrm{T}*}$-age trend, we define 
\begin{equation}
    \Delta \mathrm{V}_{\mathrm{T}*}{_\mathrm{j}} = \overline{\mathrm{V}{_\mathrm{T}*}}{_\mathrm{i}} - \overline{\mathrm{V}{_\mathrm{T}*}{_\rho}} \label{eqdeltavt} 
\end{equation}
where $\boldsymbol{\overline{V{_\mathrm{T}*}}{_j}}$ denotes the j-th substructure's mean value in the $\boldsymbol{\mathrm{V}_\mathrm{T}*}$ space and $\boldsymbol{\overline{V{_\mathrm{T}*}}{_\rho}}$ is the mean value of the $\rho$ Oph substructure in the same space. We make note that 6 Myr is assumed as the maximum age of $\rho$ Oph, but $\overline{V{_\mathrm{T}*}{_\rho}}$ would be weighted by the full range of ages. \cite{esplin2020} shown that the ages in $\rho$ Oph range from 2 to 6 Myr. Hence, we adopt 4$\pm$1 Myr as the average age of $\rho$ Oph and use this value for both plots in Figure \ref{figR.11}. In the second and fourth columns of Table \ref{Table3}, we show $\Delta \mathrm{V}_{\mathrm{T}*}{_\mathrm{j}}$ and the age estimations (with their respective upper and lower bounds) for each substructure. In upper panel of Figure \ref{figR.11} we show the $\Delta \mathrm{V}_{\mathrm{T}*}{_\mathrm{j}}$-age correlation with error bars denoting the upper and lower bounds reported in the age estimation. 

If we assume a common linear drift rate, we are able to constrain the stellar populations in USco to be $\sim$ 3.4$\pm$1 Myr older for each km s$^{-1}$ of separation with respect to to $\rho$ Oph in the $\boldsymbol{\mathrm{V}_\mathrm{T}*}$ space. However, note that there is an abrupt age difference between $\delta$ Sco and $\alpha$ Sco, despite their similar kinematic properties. There is a physical explanation to this in our picture: we expect that the parent molecular cloud of the substructures had similar kinematic properties than $\alpha$ Sco, until the first SNe in $\alpha$ Sco substructure induced a successive acceleration of their parent molecular cloud (as this is the main kinematic prediction of the triggered star formation scenario) towards the south west or en the direction of $\rho$ Oph in the $\boldsymbol{\mathrm{V}_\mathrm{T}*}$ space. Therefore, we take this as a strong suggestion of triggered star formation as the origin of the substructures younger than $\alpha$ Sco. We will discuss this with more detail in Section \ref{5.3}.

\subsubsection{Age-density correlation}\label{4.3.5}

We define the core volume of each substructure and the core mass enclosed by this volume as:
\begin{gather}
    \mathrm{V}_{\mathrm{core}} = \frac{4\pi\epsilon_{\mathrm{min}}^3}{3} \label{eqVcore} \\
    \mathrm{M}_{\mathrm{core}} = MinPts \times 0.4 M_\odot = 12.4 M_\odot
\end{gather}
This definitions are based on the core definition in the context of the OPTICS algorithm (see Section \ref{3.2.1} and do not take into account the irregularities in the spatial distribution. M$_{core}$ is obtained by converting the mean absolute G mangnitude in the USco KG to a mean mass of 0.4 M$_\odot$ based on the estimates from \cite{Pecaut2013SpT}.
Lower panel in Figure \ref{figR.11} shows the resultant empirical expansion law that correlates the density with the age of these substructures. This observational evidence suggests that each substructure expands with time at a logarithmic rate and may explain the presence of Diffuse populations as the expanded outer regions of the localized substructures found, rather than being in-situ formed populations or the result of the expansion of a unique global structure. The expansion rate probably depends on the specific formation conditions or mechanism for each structure, however, a global trend seems to be satisfied. Al thought this important result escapes the aims of this work, it deserves further investigation and further testing in other OB associations.

\begin{table}
\centering
\begin{tabular}{llll}
    \hline
    \hline
    Substructure   & $\Delta \mathrm{V}_\mathrm{T}*$ [km s$^{-1}$] & V$_{\mathrm{core}}$ [pc$^{3}$] & Age$_{\mathrm{phot}}$ [Myr]  \\
    \hline 
    $\rho$ Oph     & 0    & 6.54     & 4$\pm1$    \\
                   &      &          & 6 (max)   \\
    $\nu$ Sco      & 0.60 & 19.16    & 7$\pm1$         \\
    $\beta$ Sco    & 1.22 & 45.83    & 8$\pm1$         \\
    $\omega$ Sco   & 1.22 & 52.99    & 8$\pm1$         \\
    $\delta$ Sco   & 2.32 & 53.67    & 9$\pm2$        \\
    $\alpha$ Sco   & 2.42 & 87.11    & 14$\pm2$        \\
    $\pi$ Sco      & 4.52 & 448.92   & 20$\pm2$        \\
    \hline
    \end{tabular}
    \caption{V$_{\mathrm{core}}$ based on the $\epsilon_{\mathrm{min}}$ parameter, $\Delta \mathrm{V}_{\mathrm{T}*}$ calculated from Eq.\eqref{eqdeltavt} and Age estimated from the CMD in Figure \ref{figR.10}, via PARSEC models and the average of the extinction estimates, with the statistic errors. As discussed in the text, the intrinsic ages are also affected by a systematic error of $pm$1 Myr.}
    \label{Table3}
\end{table}

\subsection{General properties and massive content}\label{4.4}

\begin{table*}
    \centering
    \begin{tabular}{llllllll}
    \hline
    \hline
Source                  & SpT$^{\mathrm{A}}$& Mass$^{\mathrm{A}}$   & Age$^{\mathrm{A}}$& Age$^{\mathrm{B}}$     & Substructure  & Age$^{\mathrm{C}}$    \\
Designation$_\star$     &                 & [M$_\odot$]         & [Myr]           & [Myr]                &               & [Myr]      \\
\hline
$\tau$ Sco$_\star$      & B0V             & 15$\pm$0.1          & 5.7$\pm$1       & 1                    & $\alpha$ Sco* & -           \\
$\delta$ Sco$_\star$    & B0.2IV          & 13$^\mathrm{D}$     & 4.8$\pm$0.2     & 7                    & $\delta$ Sco  & 9$\pm$2    \\
$\beta^1$ Sco$_\star$   & B0.5V           & 12.5$\pm$0.6        & 13.8$\pm$0.4    & 7                    & $\beta$ Sco   & 8$\pm$1    \\
$\sigma$ Sco$_\star$    & B1III           & 18.7$^\mathrm{E}$   & 8$\pm$0.2       & 8                    & $\alpha$ Sco  & 14$\pm$2   \\     
$\omega^1$ Sco$_\star$  & B1V             & 11.1$\pm$0.9        & 10$\pm$5        & 1                    & $\omega$ Sco  & 8$\pm$1    \\  
$\pi$ Sco$_\star$       & B1V+B2V**       & 12.5$\pm$0.6        & 15.3$\pm$0.6    & 10                   & $\pi$ Sco     & 20$\pm$2   \\         
1 Sco$_\star$           & B1.5Vn          & 8.3$\pm$0.2         & 10.3$\pm$5.3    & 5                    & $\alpha$ Sco* & -           \\
$\nu$ Sco$_\star$       & B2IV            & 9.2$\pm$0.3         & 21$\pm$2.4      & 23                   & $\nu$ Sco     & 7$\pm$1    \\  
$\beta^2$ Sco$_\star$   & B2V             & 7.3$\pm$0.1         & 2.3$\pm$2.2     & $<$1                 & $\beta$ Sco   & 8$\pm$1    \\
$\rho$ Sco$_\star$      & B2IV-V          & 8.1$\pm$0.1         & 20.5$\pm$3.2    & 18                   & UCL           & -           \\ 
13 Sco$_\star$          & B2V             & 7.8$\pm$0.1         & 2.9$\pm$1.8     & 15                   & UCL           & -           \\
$\alpha$ Sco$_\star$    & M1.5Iab-b$^G$   & 11-15$^\mathrm{F}$  & 17.1$\pm$0.6    & 15$^F$               & $\alpha$ Sco  & 14$\pm$2   \\
    \hline
    \end{tabular}
    \caption{Rows: the Turn-Off sample plus $\alpha$ Sco$_\star$. Columns: 1. Designation for each star. 2. Spectral type and class. 3. Mass for the V stars is obtained from A. References are specified for the other luminosity classes. 4. Age from B. 5. Age from C. 6. Age assigned by correspondence with a substructure. 7. Parent structure. Given the mass and age from B and D, we are not able to associate $\tau$ Sco$_\star$ to any substructure. References: A: \protect\citep{massage2011runaway}, B: \protect\citep{PecautSco-cen2016},  C: this work (by association with the substructures), D: \protect\citep{deltasco}, E: \protect\citep{sigmaScoNorth2007}, F: \protect\citep{Antares2013}, *: potential coronae member. **: $\pi$ Sco$_\star$ is a spectroscopic binary comprised of a pair of stars with the shown spectral types \protect\citep{piSco1,piSco2}}.
    \label{Tabla4}
\end{table*}

We devote this section to summarize some important properties of the substructures discussed in section \ref{4.3} that we consider relevant for the discussion in Section \ref{5}. We also estimate which stars from the Turn-Off sample belong to each substructure, justifying their names. We constrain the age of these stars based on \cite{ekstromgrids} grids with rotation. In addition to the latter, this classification is also important to estimate the number of SNe events hosted by each substructure. To define if a star belongs to a given substructure, we consider mainly the spatial and celestial positions from Figure \ref{figR.7} (lower panels), and Figure \ref{figR.8}). However, there are several cases where is necessary to account for other physical considerations and some where is not possible to unambiguously relate a massive star to any substructure in particular. The results are summarized in Table , where the columns 7 and 8 constitute our age and upper mass-limit estimations.

\subsubsection{$\rho$ Oph: age cornerstone}\label{4.4.1}

Our substructure's intrinsic age estimations are strongly based on the reported maximum age of 6 Myr for $\rho$ Oph \citep{Canovas2019,esplin2020}. Hence, in section \ref{4.3}, we chose conservative $\xi$ limits for the $\rho$ Oph substructure, as we aimed to recover a confident sample, rather than an in depth survey of this well known region. We remark again that $\rho$ Oph is composed by several smaller populations with ages between 0 \citep{rhoOph0myr} and 6 \citep{esplin2020}] Myr in a complex structure of stars and molecular clouds. We will revisit these properties in our proposed star formation scenario for USco in section \ref{5}. 

\subsubsection{$\delta$, $\beta$, $\omega$ and $\nu$ Scorpii clusters}\label{4.4.2}

In previous sections we found that these substructures show a correlation in age-velocity, age-position and age-density, while they are also close in each of those spaces. We take the age-position correlation as the primary evidence of that an external triggering event induced the formation of these structures, while the other correlations independently support this scenario (see footnote 1).

$\nu$ Sco is named after the star Jabbah ($\nu$ Scorpii$_\star$), part of a hierarchical system with 7 components \citep{multiplestarcatalog2018tokovinin}. $\nu$ Sco and $\rho$ Oph barycenters are 13 pc away in the Z direction, while their dispersion in that direction is about $\sim$ 4.5 pc, however, their kinematic properties and positions in the CMD are very similar, suggesting a common mechanism for their formation and constraining the expected position of a potential triggering event towards the east. Moreover, they are both close to prominent molecular clouds located westwards (see Lynds clouds in Figure \ref{figR.8}). $\beta$ Sco is named after the star Acrab ($\beta^1$ Sco$_\star$). We recognize that there is a large difference between the distance to the cluster and the star. However, orbital motions due to its binary nature certainly affect the parallax measurements, which are from Hipparcos. This can be seen from the $\boldsymbol{\mathrm{V}_\mathrm{T}}$ positions of the pair, which locates them outside the USco KG. Hence, we argue that the distance of the pair is very likely to be underestimated due to its orbital motions and we expect it belong to the $\beta$ Sco substructure. $\omega$ Sco is named after $\omega^1$ Sco$_\star$ which is expected to be a single star \citep{Pecaut2012} and has spatial a celestial position within the boundaries of the respective structure. We note a high RUWE value ($\sim$ 2.4) for this star, suggesting that is probable that has a unresolved secondary. If verified, all the most massive stars in this substructures would be actually binaries or hierarchical systems.  $\delta$ Sco in named after Dschubba ($\delta$ Sco$\star$). We base the relation between these systems mainly on their celestial position and kinematics. 

\subsubsection{$\pi$ Scorpii: the missing link}\label{4.4.3}

$\pi$ Sco is named after the system Fang ($\pi$ Sco$_\star$), a spectroscopic binary comprised of a pair of stars with the spectral types B1V and B2V \citep{piSco1,piSco2}, or $\sim$ 12 and $\sim$ 7 solar masses. There is some ambiguity regarding the actual position in space of this system given that: the Gaia eDR3 distance to this binary star is $\sim$ 70 pc lower than the Hipparcos distance, but the Gaia eDR3 parallax error is larger that its Hipparcos counterpart, however, the resultant parallax over error value is bigger in the Gaia eDR3 catalog, $\sim$ 14$\%$, with respect to the Hipparcos value of $\sim$ 8.7$\%$. On the other hand, all the members that give their name to the substructures have extinction values A$_V$ remarkably similar to the mean A$_V$ of the substructures, which makes sense if they are embedded into the same clouds as their parent substructure. $\pi$ Sco$_\star$ is no exception to this with A$_V$ $=$ 0.2 mag, while A$_V$ for $\pi$ Sco $=$ 0.23 mag. At $\pi$ Sco$_\star$ sky position, if it were far behind the respective substructure we would expect it to have an extinction notably larger than 0.23 mag, as it is the case for 1 Sco with A$_V$ $=$ 0.46 mag. After considering this, we assign this system as a member of the respective substructure. However, this is an important caveat which needs further study. A precise estimation of fundamental parameters of Fang escapes our aims, but the mass of the B1V component is still well above the lower mass limit of the Turn-Off sample, hence, we include it as a massive member of $\pi$ Sco and a potential SNe progenitor. 

Remarkably, $\pi$ Sco overlaps with the region of high probability to be the origin of a SNe event $\sim$ -1,5 Myr ago in the outskirts of both USco, as reported by \cite{SNnature16}. This region is centered at (X,Y,Z) $=$ (83,-25,41) pc while $\pi$ Sco is located at (X,Y,Z) $=$ (101,-19, 39) pc (see Figure \ref{figR.7}). However, the study of \cite{SNzetaOph2020} on the runaway star $\zeta$ Oph \citep{zetaOphhoogerwerf2000}, provides a correction on the time and present-day location of the event, 1.78$\pm$0.21 Myr ago and (X,Y,Z) $=$ (99,-34,25) pc (see \cite{SNzetaOph2020}, its Table 4), respectively, locating it even closer at no more than 5 pc from $\pi$ Sco and. We take this as strong evidence of $\pi$ Sco being the parent structure of both $\zeta$ Ophiuchi and the progenitor of the SNe event that expelled it. This intrinsic relation provides valuable information for both the objects and the structure. For example, there is a large uncertainty on the mass of $\zeta$ Oph, 13$^{+10}_{-6}$, and the mass of its progenitor strongly depends on the age of the structure that hosted it. Indeed, \cite{SNzetaOph2020} estimated the mass of this progenitor as 17$\pm$2 M$_\odot$ based on an age of 12-15 Myr from the age map in \cite{PecautSco-cen2016}. We stress here that this age map is a sky projection, so that it does not excludes the possibility of having an older substructure in foreground, as we suggest for $\pi$ Sco. Hence, our age for $\pi$ Sco implies a new mass for the progenitor of 11.8 M$_\odot$ (see Table \ref{Table5}).

Remarkably, a Bubble in the ISM was found by \cite{robitaillebubble} in the Planck data \citep{plackdust}. \cite{robitaillebubble} suggested that this Bubble may be a $\sim$ 3 Myr old SN remnant at a distance of $\sim$ 139$\pm$10 pm, with its shell interacting with both $\rho$ Oph and Lupus clouds. While the spatial description is consistent with the observed voids in Figure \ref{figR.7}, a recent SN event in $pi$ Sco makes it more plausible for the Bubble to be related with the Void 2.

\subsubsection{$\alpha$ Scorpii moving group}\label{4.4.4}

This substructure is the largest, roughly coeval and compact structure in USco. It belongs to the $\alpha$ KG retrieved from the Kinematic Analysis. Its notable size, its relation with the $\alpha$ KG and the conclusions from section \ref{4.3.2} suggest that is largely disperse and, hence, that is the origin of many other loose members from the Diffuse populations. Given its age, we expect that stars with M $>$ 1.6 M$_\odot$ (SpT $\sim$ F0) have already entered the H-burning phase \citep{pmssiess}, suggesting a negligible number of A stars from this population in the OB sample.

$\alpha$ Sco is named after the star Antares ($\alpha$ Sco$_\star$). Antares has an age and mass of 11-15 Myr and 15$\pm$5 M$_\odot$ \citep{Pecaut2012,Antares2013}, which is consistent with the estimated age of 14 Myr and the upper mass-limit of 14.9$^{+1.8}_{1.4}$ M$_\odot$ for this substructure. Similarly, the star Alniyat ($\sigma$ Sco$_\star$) has a mass of 18.4$\pm$5.4, which is also consistent with the upper mass-limit for the substructure. Furthermore, we able to relate the system EPIC 203710387, an eclipsing binary system with contrained age from \cite{trevor2019}, to the $\alpha$ Sco substructure. According to the authors, EPIC 203710387 should have $\sim$12$\pm$3 Myr in PARSEC V1.1, which is consistent with our assigned age of $\sim$14$\pm$2 Myr with the same isochrone model.

We assign both stars to this substructure by further considering their spatial positions. Note from Figure \ref{figR.7} that is the nearest structure to these stars, so that if they do not belong to it, they must be runaways from the younger structures. We rule out the former scenario for Antares, based on its estimated age and mass. For Alniyat, the age estimates of $\sim$ 8 Myr seem to be consistent with this case, however, we also consider that the typical mass of $\beta$ Ceph stars is $\sim$ 12 M$_\odot$ and the fact that its distance seems to be systematically underestimated. Furthermore, from a canonical IMF, we expect more that one massive member for this substructure, as it doubles the stellar content of the others, despite of a large fraction of its members being already too disperse to be classified as such. Based on the latter, we argue that is more likely that the mass and age of Alniyat to have been over and sub-estimated, respectively, than it being a runaway star or have formed in situ. We classify the stars $\tau$ Sco$_\star$ and 1 Sco$_\star$ as potential coronae members of $\alpha$ Sco. However, $\tau$ Sco$_\star$ seems to be too massive and too young to have formed in this substructure and 1 Sco$_\star$ classification as such results ambiguous due to its large distance from $\alpha$ Sco.

\begin{table*}
\centering
\begin{tabular}{lllllllllllll}
    \hline
    \hline
    Structure   & N & Age          &H-M$_{\mathrm{up}}$       &He-M$_{\mathrm{up}}$       &M$_{\mathrm{up}}$      &N$_{\mathrm{SN}}$   &k                 & $\tau_1$             & M$_1$   &$\tau_2$   & M$_2$   \\
                &   &[Myr]         & [M$_\odot$]              & [M$_\odot$]               & [M$_\odot$]           &                    &                  &[Myr]            &[M$_\odot$]       &[Myr] &[M$_\odot$]       \\
    \hline 
    $\pi$ Scorpii   & 2 &20$^{+2}_{-2}$& 11.5$^{- 0.6}_{+0.7}$    & 12.2$^{-0.6}_{+0.8}$     & 12.9$^{-0.8}_{+1.1}$   & 1.7$^{+0.3}_{-0.3}$ &73.9$^{+6.2}_{-6.3}$&-11.1$^{-0.4}_{+0.1}$&32.0$^{-8}_{+20}$&-                        &-         \\
    \hline
    
                    & 2 &              &                          &                          &                        & 1.2$^{+0.3}_{-0.3}$ &63$^{+6.2}_{-5.2}$&-8.7$^{+0.4}_{-0.8}$&38.4$^{-14}_{+85}$&                     
                    &                   \\ [0.1cm]
    $\alpha$ Scorpii& 3 &14$^{+2}_{-2}$ & 14.5$^{-1.4}_{+2}$       & 15.8$^{-1.6}_{+2}$       & 14.9$^{-1.4}_{+1.8}$   & 1.7$^{+0.5}_{-0.4}$ &95.8$^{+9.5}_{-8}$&-12$^{+1}_{-0.2}$   &181$^{+60}_{-140}$&-5.8$^{-0.1}_{+0.01}$                     &22.5$^{-3.6}_{+9}$ \\ [0.1cm]
                    & 4 &              &                          &                          &                        & 2.3$^{+0.6}_{-0.5}$ &127$^{+12.5}_{-10.5}$&-8.7$^{+0.4}_{-0.8}$&38.4$^{-14}_{+85}$&-4.5$^{+0.3}_{+0.3}$ 
                    & 19.7$^{-2.8}_{+4.1}$\\
    \hline
    
    $\delta$ Scorpii& 1 & 9$^{+2}_{-2}$& 21.1$^{-3.4}_{+8.6}$     &23.2$^{-4.1}_{+9.9}$       & 13                    & $<$0.63            &22$^{+2}_{-2}$    & -                  &    -     & -        & -          \\
    \hline
    $\beta$ Scorpii & 1 & 8$^{+1}_{-1}$& 24.5$^{-3.4}_{+5.2}$     &27.4$^{-4.2}_{+5.7}$       & 12.5                  & $<$0.62            &21$^{+1.3}_{-1.2}$& -                  &     -    &  -       & -         \\
    \hline
    $\omega$ Scorpii& 1 & 8$^{+1}_{-1}$& 24.5$^{-3.4}_{+5.2}$     &27.4$^{-4.2}_{+5.7}$       & 11.1                  & $<$0.73            &21$^{+1.3}_{-1.2}$& -                  &      -   &   -      &  -        \\
    \hline
    $\nu$ Scorpii   & 1 & 7$^{+1}_{-1}$& 29.7$^{-5.2}_{+7.3}$     &33.0$^{-5.6}_{+8.2}$       & 9.2                   & $<$0.89            &20$^{+1.2}_{-0.9}$& -                  &       -  &    -     &   -       \\
    \hline
    \end{tabular}
    \caption{Rows: stellar evolution parameters for each substructure. Columns: 1. Substructure's name. 2. The number of stars with mass $>$ 7.3 M$_\odot$ 3. Substructures age with statistical uncertainty. 4. Expected upper limit for the mass of H-burning stars. 5. Expected upper limit for the mass of He-burning stars. 6. Mass of the more massive star. For $\alpha$ and $\pi$ Sco$_\star$ is the mass of Antares and $\zeta$ Ophiuchi's companion  7. N$_{SN}$ results (see Section \ref{3.3}). 8. Normalization constant (see Section \ref{3.3}) 9. Occurrence time for the first SNe retrieved. 10. Mass of progenitor at $\tau_1$. 11. 12. Mass of progenitor at $\tau_2$.}
    \label{Table5}
\end{table*}

\subsection{SNe in USco}\label{4.5}

To obtain N$_{\mathrm{SN}}$, we follow the procedure sketched in Sect \ref{3.3}. We obtained that only the substructures $\pi$ Sco and $\alpha$ Sco have possibly hosted at least 1 SNe in the past, so we focus our exposition on these substructures. The input parameters for each substructure are N, M$\tau$ and M$_{\mathrm{up}}$, where N is the number of stars with M $>$ 7.3 M$_\odot$ (SpT $<\sim$ B2V), M$\tau$ is the mass of a SNe progenitor occurred at the time $\tau$ after the structure's birth and M$_{\mathrm{up}}$ is the mass of the most massive member in the given structure. The introduction of parameters is not straightforward, for example, in section \ref{4.4} we obtained with certainty that N$=$2 for $\pi$ Sco but N may be 2, 3 or 4 for $\alpha$ Sco, so we evaluate three cases for this substructure. On the other hand, we lack precision on the mass of the most massive member for both substructures and the occurring times and progenitor mass for the SNe are very sensible to this value. We were able to solve the the M$_{\mathrm{up}}$ problem in a consistent manner for each substructure. The most massive star in $\alpha$ Sco is Antares, which is already in the He-burning phase, hence, its mass is somewhere between the upper mass for H-burning and the upper mass for He-burning, where both depend on the age $\alpha$ Sco. This way, we obtain the mass of Antares and the SNe occurring times with uncertainty $<$ 1 M$_\odot$ and 1 Myr in the full range of the uncertainty for the age of $\alpha$ Sco. For $\pi$ Sco, we use the occurring time of the SNe that expelled $\zeta$ Oph: 1.78$\pm$0.21 Myr (see Section \ref{4.4.4}); for a given age of $\pi$ Sco, we only have to find the progenitor mass of a SNe that occurred at that time. This mass depends on the age of $\pi$ Sco, so we also explored the upper and lower bounds of its age. For the remaining substructures, we obtain the normalization constant based on the upper limit-mass and estimate N$_{\mathrm{SN}}$ based on the masses from Table \ref{Tabla4}. For the remaining substructures, we obtain N$_{\mathrm{SN}} <$ 1. Remarkably, this implies that the present day mass function is identical to the IMF in these substructures. However, we note that the upper mass values notably differ from the measured values, most dramatically in the case of the $\nu$ substructure, in contrast with the observed on the previously discussed substructures where the difference was minimal. The latter may be consequence of the structures being smaller or evidence of a more fundamental relation between the IMF and the respective formation mechanism. This is an important result that deserves further investigation and we will revisit this in future work.

Via this procedure, we obtain that at least 3 and up to 4 SNe have occurred in USco. 2 hosted by $\pi$ Sco $\sim$ 10 and 1.78 Myr ago and 1 or 2 hosted by $\alpha$ Sco with occurring times depending on N in Table \ref{Table5}. Note that the 3 values of N and age for $\alpha$ Sco yield a SNe $\sim$ 10$\pm$2 Myr ago, allowing us to conclude that a SNe certainly happened around this time. However, the mass of the progenitor of this SNe largely depends on the assumed age for $\alpha$ Sco. For N $=$ 3, 4, we obtain a second event $\sim$ 5-6 Myr ago. Hence, we obtain that 2 SNe occurred $\sim$ 10$\pm$2 Myr ago, another $\sim$ 5 Myr ago and a more recent one, the one that would have expelled $\zeta$ Oph, 1.78 Myr ago. We base mainly on this, the internal structure of USco and the age gradients observed to suggest a star formation scenario for USco in section \ref{5.3}.

\section{Discussion}\label{5}

\subsection{Robust substructure retrieval}

It is straightforward to conclude from Sect \ref{4.2} that the spatial distribution of the KGs is genuinely complex, justifying the need of the Spatial Analysis (see Section \ref{3.2}) to extract more information from this region. Moreover, evidence of substructure in USco with Gaia DR2 \citep{kerr} and Gaia eDR3 \citep{Squicciarini} has already been found. However, there are important differences between our results and those of these works due to the different procedures and the improved precision of Gaia eDR3. For example, \cite{kerr} recovers 7 substructures in USco. However, the authors set the prior of spatial substructures in USco to be clustered in the transverse velocity space, which is physically expected, but, given how close the substructures retrieved are in this kinematic space and that the typical errors in Gaia DR2 \citep{Gaiadr2} increase by a factor of 3 to 4 in proper motion and 30-50\% in parallax, we argue that this is not a good empiric assumption in the particular case of USco (see Section \ref{1}) and Section \ref{3.2.1}). We have also shown how unresolved binaries would introduce a strong bias into this assumption in Sect \ref{4.3.2}. \cite{Squicciarini} recovers 8 substructures in USco by tracing back in time the position of the sources in USco and applying a k-means clustering. The authors validate the physical relevance of some of the substructures retrieved via a visual comparison with the results from \cite{kerr}, while recognize that 2-3 of them require further investigation. 

We were able to unambiguously relate substructures 17-E, 17-G, 17-I and 13 from \cite{kerr} to $\nu$ Sco, $\beta$ Sco, $\rho$ Oph and $\pi$ Sco, respectively. Our Spatial Analysis suggests that their substructure 17-H is actually composed by $\omega$ and $\delta$ Sco, while substructures 17-B and 17-C are contained within the larger $\alpha$ Sco. Finally, 17-A and 17-F do not have any counterpart in our work, probably due to the reasons discussed previously in this section. On the other hand, substructures 1, 2 and 6 from \cite{Squicciarini} have share sky positions and tangential velocities with $\rho$ Oph, $\nu$ Sco and $\beta$ Sco. Substructures 3 and 4 have similar sky positions to $\omega$ Sco and $\delta$ Sco, but notably different kinematics. Correspondence between substructures 3, 7 and 8 with any of our work is ambiguous and hard to conclude since we do not have access to this data. From this comparison, we show that we are able to significantly extend the membership and improve the characterization of the already known structures $\nu$ Sco, $\beta$ Sco, and $\pi$ Sco, while our analysis clearly distinguishes the presence, spatial and kinematic clustering of the structures $\alpha$ Sco, $\omega$ Sco and $\delta$ Sco via the sum of independent methods in a multidimensional approach. We are also able to relate for the first time these substructures with the massive stars after which they are named, and provide quantitative evidence (the $\epsilon$ and $\xi$ parameters) of their clustered nature. 

\subsection{The age of USco}\label{5.1}

USco has been historically treated as a coeval population younger than the other groups in Sco-Cen. However, in section \ref{4.3.1} we show that USco is far from coeval in the Myr scale, as there are stellar objects with ages ranging from 0 to $\sim$ 20 Myr. Moreover, we not only observe a large age dispersion, but also show that several distinct populations are responsible for this spread. We recover a mean age of $\sim$ 9.4 Myr based on the substructures shown in Figure \ref{figR.10}. Our results are in agreement with previous recent estimations of $\sim$ 10 Myr as the global age of USco with a broad variety of methods \citeg{Pecaut2012,PecautSco-cen2016,Sullivan2021,kerr}, with the exception of the estimated age of 4.5$\pm$0.1 Myr for the clustered populations and 8.2$\pm$0.1 Myr for the diffuse populations by \cite{Squicciarini}. However, we note that relative ages between common structures are in agreement. We restate the previous conclusions of USco being younger than the other groups in Sco-Cen as it having a large fraction of younger populations than the other groups, biasing its average age towards the 10 Myr value, but its older structures, $\alpha$ and $\pi$ Scorpii, are roughly coeval with the other groups in Sco-Cen. Moreover, the presence of these substructures and the massive stars associated with each also explains why these massive members do not span a clear turn-off sequence \citep{Pecaut2012,PecautSco-cen2016}. This work also constrains the age of several of the most massive members in USco, which yield differences $>$ 10 Myr between their ages. Of course, this age difference extends to the mid and low mass populations, hence, future works would have to consider the positions, kinematics, dynamics and ages of these substructures, observationally constrained in this work, for the full mass range in the USco region. This improvement in the resolution of these properties makes USco an even more valuable region as it spans multiple young populations with now resolvable ages that will make it possible to further constrain processes of planetary formation, early stellar evolution and star formation.

\subsection{USco: home of near Earth supernovae}\label{5.2}

In section \ref{4.5} we have shown that is very likely that 3 to 4 SNe events have occurred in $\alpha$ Sco: $\sim$ 10$\pm$2 (SN$_\mathrm{A1}$) and 5$\pm$1(SN$_\mathrm{B}$) Myr ago, and in $\pi$ Sco: $\sim$ 10$\pm$2 (SN$_\mathrm{A2}$) and 2(SN$_\mathrm{C}$) Myr ago. Here we briefly discuss these events in relation with some previous work 
There is a copious amount of evidence for $^{60}Fe$ ejecta from recent $<$ 10 Myr SNe events to have reached the Earth (see \cite{nearEarthSNe10myr}, its Figure 1 and references therein). In this regard, the candidates to have hosted these events are Sco-Cen and the Tucanae-Horologium (Tuc-Hor) associations \citep{benitez2002,fuchs2006,tucanaesupernovae}. Although some models favor Tuc-Hor when trying to explain the recent $\sim$ 2.2 Myr $^{60}Fe$ peak \citeg{Fry2016100pc,Miller100pc}, only one SNe event is speculated to have occurred in this association, making it necessary to look somewhere else when trying to explain the more ancient $\sim$ 7 Myr peak in the $^{60}Fe$ income.

\cite{60feejecta} concludes a realistic signal can be produced by a SN in the distance range of 10 to 200 pc and requires two SNe to explain the observed $^{60}Fe$ peaks. Based on this we suggest a contribution from the expected SN$_{A1}$ and SN$_{A2}$ events to the 7 Myr peak. This is also consistent with the results of \cite{nearearthSN35myr}, that reports a peak in the SNe events in the solar neighborhood 10 Myr ago. The recent 2.2 Myr peak may be explained by the SN$_B$ event. For the ejecta from the recent SN$_C$ we can conceive 2 possibilities: 1. It is about to reach the Earth as it was similarly concluded by \cite{SNzetaOph2020}, or 2. If we relax the uncertainty in the occurrence time of the SN$_C$, justified by the age estimation of the \cite{robitaillebubble} Bubble (3 Myr), the SN$_C$ may have also contributed to the 2.2 Myr peak.  
Some considerations which prevents us to go beyond suggesting this scenarios are the following: 1. Given that SN$_B$ position is highly uncertain, the $\alpha$ Sco distance of $\sim$ 160 pc may be too far for its SN$_{A2}$ ejecta to reach us \citeg{Fry2016100pc,Miller100pc}. 2. The $^{60}Fe$ peaks may be the result of a complex chain of SNe events, instead of just one per peak, as it was found by \cite{Wallner2016}. 3. The expected distances and reaching times rely on assumptions like a uniform ISM and a given velocity of the solar system with respect to the SNe shock front \cite{60feejecta}. 

On the other hand, \cite{Forbes} investigated the origin of the observed amount of $^{26}$Al radionuclide in the Ophiuchus clouds and leaves the age of USco as a free parameter to constrain its source from being stellar winds and SNe in different proportions. To this date only 1$\pm$1 SNe was expected to have ocurred in USco \citep{deGeusHIshells} given it was assumed to be 5 Myr when this estimation was last performed. In this work show that at least 3 and up to 4 SNe have occurred in USco at different times in the last $\sim$ 12 Myr. In this sense, our results partially constrain the source type of $^{26}$Al, since we update the age of USco and its internal stellar populations and, consequentially, the amount of expected SNe in the region. In this regard, a recalculation of energy input of the USco-Shell considering contribution of the stellar winds of the missing massive stars and the respective SNe events becomes necessary. This, together with the $^{26}$Al constrains, would place tighter constrains on how many SNe have actually gone off in USco. However, this matter escapes the aims of our work.

We report the discovery of $\pi$ Sco being the structure that hosted the SNe event that expelled $\zeta$ Oph \citep{SNzetaOph2020}, which has been also suggested to be one of the two nearest SNe events to Earth \citep{SNnature16}. 

\subsection{Star formation history}\label{5.3}

A plausible scenario for the star formation history (SFH) of Sco-Cen was proposed by \cite{kerr}. However, the SFH of USco has been subject of debate for decades \citeg{deGeusHIshells,preibisch1999,Preibisch08,Pecaut2012,PecautSco-cen2016,Damiani2019,Luhman2020,Squicciarini}. Here we suggest a SFH with unprecedented time-resolution based on the uncovered spatial, kinematic and age substructure of USco that allowed us to recover potential past SNe events in this region and find evidence of the kinematic imprints of the progenitors on the young stellar populations. 
Our proposed scenario is focused on explaining how the substructures $\rho$ Oph, $\nu$ Sco, $\beta$ Sco, $\omega$ Sco, and $\delta$ Sco were formed. We suggest that the substructures retrieved are the result of a complex chaining of triggering events associated with the expected SNe events and the stellar winds of their massive progenitors. In our scenario, expanding shells powered by the progenitor winds and remnants of the SN$_\mathrm{A1}$ and SN$_\mathrm{A1}$ triggered the compression of the molecular clouds between them at their encounter fronts $\sim$ 10 Myr ago, whose collapse had been already stopped or slowed down by the stellar winds of their massive progenitors. These events would have induced the formation of the $\delta$ Sco, $\beta$ Sco, $\omega$ Sco, $\nu$ Sco substructures, leaving as evidence the age gradients observed today in Figures \ref{figR.9},\ref{figR.10} and \ref{figR.11}. SN$_\mathrm{B}$ and SN$_\mathrm{C}$ events would have further triggered star formation in the $\rho$ Oph molecular cloud complex. We base these conclusions on the following:
\begin{itemize}
  \let\labelitemi\labelitemii
    \item The age-position gradient and small size of the substructures with age $<$ 10 Myr (see Figure \ref{figR.7} and Table \ref{Table2}) plus their age-velocity and age-density correlations (see Section \ref{4.3.4}) suggest an external triggering event, as their parent molecular cloud probably did not have enough mass to collapse at the time the more massive $\alpha$ Sco was formed.
    \item In Figure \ref{figR.7}, we observe two regions (Void 1 and Void 2) with an apparent lack of stars or dense regions. These voids are well fit by ellipsoids with semi axes in the outskirts of $\pi$ Sco and $\alpha$ Sco. The remaining substructures lay between these voids and separate them. This ellipsoidal geometry is not likely to be coincidental and it also correlates with the age-position gradient. Hence, we deduce that the lack of stars in these regions is the result of a true physical process, i.e., stellar winds from the progenitors of SN$_\mathrm{A1}$ and SN$_\mathrm{A2}$ altering the dynamics and geometry of the ancient molecular clouds. 
    \item One these above mentioned progenitors was hosted by $\alpha$ Sco and the other by $\pi$ Sco on the opposite sides of the younger substructures, which are located at the edges of these voids.
    \item The occuring time of SN$_\mathrm{A1}$ and SN$_\mathrm{A2}$ coincides with the maximum age of $\delta$ Sco.
    \item The global age of $\rho$ Oph complex can be explained by the above mentioned scenario, but its internal ages do not show any clear directionality. Recently, \cite{rhoOphSFH} found that more evolved stars are farther from the denser clouds and suggest that the shape of the cloud B in its Figure 1 is due to the shocks of the stars belonging to our $\alpha$ Sco substructure. This morphology, we argue, may also be explained by the effect of SN$_\mathrm{B}$ and SN$_\mathrm{C}$ events.
    \item The occurring time of the SN$_\mathrm{B}$ is remarkably consistent with the maximum age of $\rho$ Oph complex, suggesting that it could have reactivated or definitely induced star formation in the complex.
    \item Our scenario is consistent with the predicted star formation mechanism in Sco-Cen by \cite{surroundandsquash2017}, based on different observations before Gaia-DR2. \cite{surroundandsquash2017} concluded that similar a scenario is the predominant mechanism for star formation in Sco-Cen and possibly other OB associations and questioned the independence of the star formation in Lupus I and $\rho$ Oph clouds. Hence, the SN$_\mathrm{C}$ may be responsible for the most recent star formation activity in the $\rho$ Oph and Lupus clouds, in agreement also with the scenario suggested by \cite{lupussquezed}.
    \item A similar formation scenario has been proposed in specific for the Lupus I clouds by \cite{lupusshells}, where the recent star formation in this region would be the result of the interaction between the USco Shell and the stellar winds from B stars in UCL. In this respect, we also observe that the Void 2 limits with the Lupus I clouds, extending our scenario to that region and supporting their conclusions. Moreover, as we have mentioned, we expect that the progenitor of the expected SNe events would also have contributed to the Voids and the arrangement of the $<$ 10 Myr substructures' parent clouds.
    \item The age and location of the clouds in USco, USco-Shell (5.7 Myr) and USco-Loop (1 Myr) \citep{poppelHIloopsshells}, coincide remarkably well with both the occurring times and locations of the SN$_\mathrm{B}$ and SN$_\mathrm{C}$ respectively, implying that these structures are the remnants of these SNe events.
    \item The $\sim$ 3 Myr old bubble observed by \cite{robitaillebubble} overlapping with the $\pi$ Sco age and position, which the authors suggest can be originated by a SN remnant with its shells interacting with both $\rho$ Oph and Lupus clouds.
\end{itemize}

Additionally each SN-triggering event is supported by or explains several independent observations. SN$_\mathrm{A1}$ and SN$_\mathrm{A2}$: the voids and geometry observed in Figure \ref{figR.7}, the existence of the age, density and kinematic gradients of the structures located between $\pi$ Sco and $\alpha$ Sco.  SN$_\mathrm{B}$: the maximum of the $\rho$ Oph complex, the Lupus I suggested formation scenario and the USco-Shell. SN$_\mathrm{C}$: the $\zeta$ Ophiuchi runaway star. All of these observations point to and are unified by the proposed scenario.

\section{Summary and conclusions} \label{5}

In the following, we summarize the main results of this work:

1. We retrieve a sample of 3661 candidate members for Upper Scorpius (USco) from the Gaia-eDR3 catalog, with an estimated contamination of $\sim$ 9\% from Galactic field, and successfully validate it with the previous work from \citep{Damiani2019,Luhman2022}. We also compile an astrometrically cleaner sample of 3004 sources with $\sim$ 6\% of contamination (see Table \ref{Table1.5}). 

2. We show that Upper Scorpius is composed by three main Kinematic Groups (KGs) with a clear age trend (see Figure \ref{figR.5}) and a complex spatial distribution (see Figure \ref{figR.6}). Remarkably, we note that single KGs are related to multiple spatial overdensties, suggesting that kinematic-only clustering spaces do are not enough to resolve all the spatial substructure in USco (and probably in other young associations).

3. To further investigate the spatial distribution of the KGs, we make use of the OPTICS algorithm \citep{OPTICS}. We apply OPTICS in the Heliocentric coordinates (X,Y,Z) space to exclude any bias due to non-negligible errors and kinematically undistinguishable stellar populations. This analysis produces robust evidence of the presence of seven clustered substructures with the suggested names: $\pi$ Scorpii (20 Myr), $\alpha$ Scorpii (14 Myr), $\delta$ Scorpii (9 Myr), $\beta$ Scorpii (8 Myr), $\omega$ Scorpii (8 Myr) and $\nu$ Scorpii (7 Myr), after their most massive member, plus the well known $\rho$ Ophiuchi substructure and a component of loose populations spanning the entire region (see Figure \ref{figR.7}). 

4. We constrain the membership, age and upper mass limit of several massive members in USco, including Antares (see Table \ref{Tabla4} and Table \ref{Table5}). We also have found that $\omega^1$ Sco is very likely to have an unresolved binary companion.

5. We observe two quasi-ellipsoidal stellar Voids which probably trace the stellar feedback from ancient massive stars (see Figure \ref{figR.7}). The denser and younger regions in USco are located at the edges of these Voids. This fact, suggests that the now-gone massive members shaped and induced the subsequent star formation in the region.

6. We constrain the occurrence of two SN events, hosted by the $\alpha$ Sco and $\pi$ Sco substructures, $\sim$ 10$\pm$2 Myr ago, a third SN hosted by $\alpha$ Sco $\sim$ 5$\pm$1 Myr ago and a fourth $\sim$ 2 Myr ago hosted by $\pi$ Sco. We identify the latter with the event that expelled the $\zeta$ Ophiuchi runaway star.

7. We observe an age-position and age-velocity correlation in the spatial substructures found in USco (see left panel in Figure \ref{figR.9}). We base on this correlations to propose an star formation inducing event coming from the east. We identify this event with the expanding shells centered in the Voids and powered by the stellar feedback of the missing massive stars that went off as SN $\sim$ 10$\pm$2 Myr ago.

8. We suggest a star formation history for USco in which 4 ancient massive stars from $\alpha$ Sco and $\pi$ Sco shaped its present structure via their stellar winds and SNe explosions. In this scenario, the younger substructures $\delta$ Sco, $\beta$ Sco, $\omega$ Sco, $\nu$ Sco and $\rho$ Oph are located between two stellar voids caused by the mentioned feedback events and they trace its influence with trends in their positions, kinematics, age and density.
 
9. This scenario is similar to that proposed by \cite{surroundandsquash2017} and we support the hypothesis of a common origin for the $\rho$ Oph complex and the Lupus I clouds \citep{lupussquezed}. Additionally, the latter would be the SN event that expelled the runaway star $\zeta$ Oph \citep{zetaOphhoogerwerf2000,SNzetaOph2020}. All these events would also be related to the observed peaks in $^{60}$Fe depositions measured on Earth \citep{recentSN10Myr}. The occurring times, masses and other parameters of interest associated to this events are summarized in Table \ref{Table5}.

10. We suggest a direct relation between the proposed SNe events in USco and other independent observations, like the $^{60}$Fe peaks observed in the Earth and the Moon in the last $\sim$ 10 Myr \citep{recentSN10Myr}, the $^{26}$Al abundance observed in USco \citep{Forbes}, the structure of the $\rho$ Ophiuchi molecular clouds \citep{rhoOphSFH}, recent star formation in the $\rho$ Ophiuchi and Lupus clouds \citep{lupusshells}, the well known USco Shell \citep{deGeusHIshells} and a recently found Bubble in USco \citep{robitaillebubble}. 

11. We provide an empirical expansion law for the substructures in USco (see lower panel in Figure \ref{figR.11}). We base on this law we suggest that the loose populations of USco, and probably the other groups in Sco-Cen, have origin in the core regions rather than being formed in situ. We make note that the assumption that the expansion rate is global does not takes into account key factors like the formation mechanism for each substructure and the different initial conditions. The universality of this law is to be tested in future work.

\section*{ACKNOWLEDGEMENTS} 

 Authors acknowledge the referee Eric Mamajek for the very insightful comments that helped to improve this work. Authors also acknowledge Kevin Luhman for comments and discussion on the membership validation of our census, Ralph Neuh\"auser for comments and discussion about our scenario in which $\pi$ Sco hosted the recent supernova event that expelled the star $\zeta$ Ophiuchi and the update of the mass of this supernova progenitor, and Manuela Zoccali for general comments about the Kinematic Groups found in Section \ref{4.2}. GBM also acknowledges David Sánchez Elizondo for discussions on nature, phenomena and causality, and support from Central American and Caribbean Bridge in Astrophysics (Cenca Bridge), Astrofísica Centroamericana y del Caribe (Alpha-Cen) and Centro Latinoamericano de Física, subsede Centroamérica (CLAF-CA). JC acknowledges support from the Agencia Nacional de Investigación y Desarrollo (ANID), via Proyecto FONDECyT Regular 1191366, and by the ANID BASAL projects CATA-Puente ACE210002 and CATA2-FB210003. 

\section*{DATA AVAILABILITY}

This work presents results from the European Space Agency (ESA) space mission Gaia. Gaia data are being processed by the Gaia Data Processing and Analysis Consortium (DPAC). Funding for the DPAC is provided by national institutions, in particular the institutions participating in the Gaia MultiLateral Agreement (MLA). The Gaia mission website is https://www.cosmos.esa.int/gaia. The Gaia archive website is https://archives.esac.esa.int/gaia. The data from Table \ref{Table1.5} will be available via the VizieR-Strasburg service.



\bibliographystyle{mnras}
\bibliography{bib.bib} 


\clearpage

\appendix

\section{Coordinate and space transformations}\label{A}

A kinematic group (KG) its defined as a group of stars with a similar space velocity, so we can identify them with over densities in the (U,V,W) velocity space. Moreover, it is useful to define a KG's barycenter\footnote{Formally, the center of mass instead of the barycenter should be obtained, but the barycenter approximation does no change the results in a significant way (\citealt{perryman1998}, \citealt{lodieuhyadess}).} such that a barycenter motion (U, V, W)$_{bar}$ can be calculated and taken as the true KG's motion. In order to do this this, we closely follow the procedure of \cite{perryman1998} (see its section 5.1 for details). For clarity, an explanation of the differences in our procedure and the main common steps is sketched bellow.

We take the equatorial coordinates of the north Galactic pole in the ICRS to be 

\begin{gather}
    \alpha_G=192^\circ.859 48 \\
    \delta_G=+27^\circ.128 25 \;
\end{gather}

With this convention \citep{1997ESA}, we are able to obtain the Galactic proper motion ($\mu_{\ell}$, $\mu_b$) from the equatorial ICRS proper motion Gaia measurements ($\mu_{\alpha \star}=\mu_{\alpha} cos \delta$, $\mu_\delta$) by directly performing the rotation:

\begin{equation}
\left(
\begin{array}{c}
\mu_{\ell}  \\
\mu_{\textit{b}}  
\end{array}
\right)
= \frac{1}{\cos b}
\left(
\begin{array}{cc}
a_1 & a_2  \\
-a_2 & a_1 
\end{array} \right)
\left(
\begin{array}{c}
\mu_{\alpha \star}  \\
\mu_\delta  
\end{array} \right)
\end{equation}

With a$_1$, a$_2$ and $\cos b$ are given by

\begin{equation}
    a_1=\sin\delta_G\cos\delta-\cos\delta_G \sin\delta\cos(\alpha-\alpha_G)
\end{equation}
\begin{equation}
    a_2=\cos\delta_G\sin(\alpha-\alpha_G)
\end{equation}
\begin{equation}
    \cos b = \sqrt{a_1^2+a_2^2}
\end{equation}

See \cite{Poleski2013} for a detailed derivation. 
Then, the tangential velocity components can be calculated as:

\begin{equation}
\left(
    \begin{array}{c}
    V_\ell \\
    V_b 
    \end{array}
\right)
=
\frac{C k}{\pi}
\left(
    \begin{array}{c}
    \mu_\ell \\
    \mu_b 
    \end{array}
\right)
\label{eqVlyVb}
\end{equation}

where C$=$4.74047 km yr s$^{-1}$ is the conversion constant, k is the Doppler correction factor, that we take here as 1, and $\pi$ denotes the parallax measurement, which in our case have formal errors $<10\%$, moreover, this last allows us to obtain an estimate of the distance D to a star with the approximation D $\eqsim 1000/\pi$, where $\pi$ is given in mas.
Then, the barycentric position for each star and its velocity components in Galactic coordinates can be obtained with:

\begin{equation}
\left(
    \begin{array}{c}
    X \\
    Y \\
    Z
    \end{array}
\right) 
= 
\boldsymbol{R}
\left(
    \begin{array}{c}
    0 \\
    0 \\
    D
    \end{array}
\right)
\end{equation}

\begin{equation}
\left(
    \begin{array}{c}
    U \\
    V \\
    W
    \end{array}
\right)
=
\boldsymbol{R}
\left(
    \begin{array}{c}
    V_\ell \\
    V_b \\
    V_R
    \end{array}
\right)
\end{equation}

where $\boldsymbol{R}$ is the rotation matrix:

\begin{equation}
\boldsymbol{R}
= 
\left(
\begin{array}{ccc}
-\sin\ell & -\sin b\cos\ell & \cos b\cos\ell  \\
\cos\ell & -\sin b\sin\ell & \cos b \sin\ell  \\
0           & \cos b            & \sin b
\end{array} 
\right)
\label{eqUVW}
\end{equation}

Then, the KG's barycenter and barycenter motion can be via the distribution that models the data. We note here that the main limitation in the (U, V, W) analysis is the incompleteness of V$_\mathrm{r}$ measurements in the Gaia-eDR3 catalog. In fact, in our Main Sample only $\sim 13\%$ of the stars have V$_\mathrm{r}$ measurements (14 $\lesssim$ Gmag $\lesssim$ 7 ). Therefore, it is necessary to define a space that gives us information about the true velocity space clustering when no V$_\mathrm{r}$ is available. For this we make use of the observed tangential velocity (V$_\ell$,V$_b$) and the information obtained from the clustering for stars with V$_\mathrm{r}$ measurements by defining the $\boldsymbol{V}_{T\:exp}$ space as the expected tangential velocity as a function of sky position given a true barycenter space motion, this is:

\begin{equation}
\boldsymbol{V}_{T\:exp}
=
\left(
    \begin{array}{c}
    V_\ell \\
    V_b 
    \label{eqVtexp}
    \end{array}
\right)_{exp}
=
\boldsymbol{T^\prime}
\left(
    \begin{array}{c}
    U \\
    V \\
    W
    \end{array}
\right)_{bar}
\end{equation}

where $\boldsymbol{T^\prime}$ is:

\begin{equation}
\boldsymbol{T^\prime}
= 
\left(
\begin{array}{ccc}
-\sin\ell             & \cos\ell            & 0  \\
-\sin b\cos\ell   & -\sin b\sin\ell & \cos b 
\end{array} 
\right)
\label{eqnT}
\end{equation}

Finally we choose the transformed space $\boldsymbol V_{T*}$ to perform our clustering analysis. As Upper Scorpius is near, expands a large area in the sky and we aim to distinguish its spatial and velocity substructures, doing our analysis in this space becomes necessary, so that we can minimize projection and perspective effects.

\section{Bayesian information criteria}\label{C}

\begin{figure*}
    \centering
    \includegraphics[width=17cm]{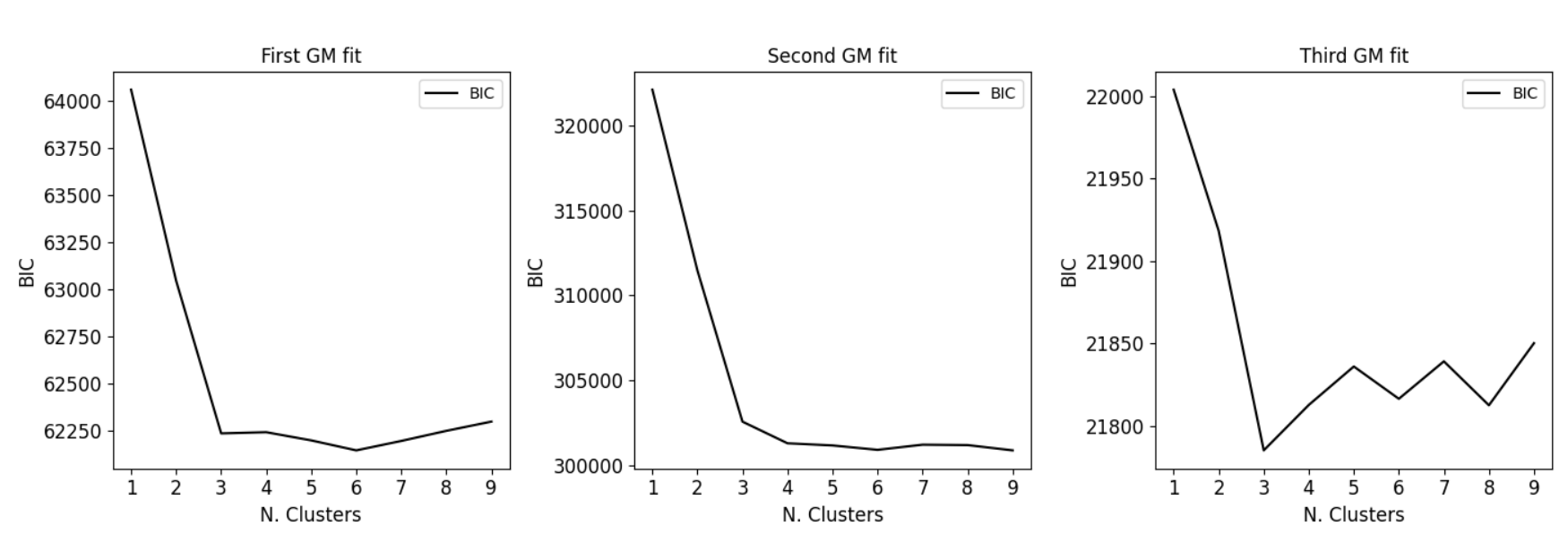}
    \caption{BIC for the First, Second and Third GM fits. 3, 4 and 3 components are chosen for these fits based on this plots.}
    \label{AppendixC1}
\end{figure*}

We base the selection of the number of components for each GM in the Bayesian Information Criteria (BIC). The BIC yields information about the goodness of fit for a range of number on components. In Fig.\ref{AppendixC1} we show the BIC for the first 9 components for each GM fit. In the First fit we chose 6 components as it is the global minimum. In the Second fit we chose the first significant local minimum, as larger number of components could yield a lower BIC, but this would compromise us with over fitting the data. In the Third fit the 3 components are chosen because 3 is the global minimum and the most simple  case.

\label{lastpage}
\end{document}